\documentclass[traditabstract]{aa} 
\usepackage{graphicx}
\usepackage{txfonts}
\begin{document}

\title{The effects of dust on the photometric parameters of decomposed disks and bulges\thanks{Appendices containing tables with corrections are only
    available at the CDS via anonymous ftp to cdsarc.u-strasbg.fr or via http://cdsarc.u-strasbg.fr/viz-bin.}}
\author{Bogdan A. Pastrav\inst{1}
          \and
          Cristina C. Popescu\inst{1}\thanks{Visiting Scientist at the Max
            Planck Institut f\"ur Kernphysik, Saupfercheckweg 1, D-69117 Heidelberg, Germany}
          \and
          Richard J. Tuffs\inst{2}
          \and
          Anne E. Sansom\inst{1}}
 \institute{Jeremiah Horrocks Institute, 
              University of Central Lancashire,
              PR1 2HE, Preston, UK\\
              \email{bapastrav@uclan.ac.uk;cpopescu@uclan.ac.uk;aesansom@uclan.ac.uk}
         \and
              Max Planck Institut f\"ur Kernphysik, Saupfercheckweg 1, D-69117
              Heidelberg, Germany\\
              \email{Richard.Tuffs@mpi-hd.mpg.de}
            }
 \date{Received / Accepted}
\abstract{
We present results of a study to quantify the effects of dust on the derived
photometric parameters of disk and bulges obtained from bulge-disk
decomposition: scale-length, effective radius, S\'{e}rsic index, disk 
axis-ratio, and bulge-to-disk ratio. The dust induced changes in these
parameters were obtained by fitting
simulated images of composite systems (containing a disk and a bulge) produced
using radiative transfer calculations. The simulations were fitted with the
GALFIT 3.0.2 data analysis algorithm. Fits were done with both a combination of
an exponential plus a variable-index S\'{e}rsic function as well as with a
combination of two variable-index S\'{e}rsic functions. We find that dust
is biasing the derived exponential scale-length of decomposed disks towards 
smaller values than would be otherwise derived if the galaxy were to have no
bulge. Similarly, the derived bulge-to-disk ratio is biased towards smaller
values. However, the derived axis-ratio of the disk is not changed in the 
decomposition process. The derived effective radius of decomposed disks of
systems having exponential bulges is 
found to be less affected by dust when fits are done with two variable-index 
S\'{e}rsic functions. For the same type of fits dust is found to bias the value
of the derived effective radius of decomposed disks towards lower values
for systems having de Vaucouleurs bulges. All 
corrections derived in this paper are made available in electronic form.}

 \keywords{galaxies: spiral -- galaxies: bulges -- galaxies: photometry -- galaxies: structure -- ISM: dust, extinction -- radiative transfer}
 \maketitle

\section{Introduction}

Spiral galaxies are complex systems containing two primary, physically 
distinct morphological components: a disk and a bulge. The bulge is a
predominantly pressure-supported spheroidal component containing old stellar
populations. Being pressure supported, there can be no substantial 
cold interstellar medium associated with the
spheroid.  Consequently, it is believed that there is no dust associated with 
this component. Conversely, the disk is a flat, rotationally-supported 
component containing young, intermediate-age and old stellar populations, with 
star-formation activity mainly occurring in a system of spiral arms. Unlike the
bulge, the disk is associated with a cold interstellar medium, and contains 
large amounts of dust. The dust in the disk has the effect of attenuating the stellar light from both the disk and the bulge (e.g. Tuffs et al. 2004, Driver et al. 2007).

Although the bimodal structure of spiral galaxies has long been known, the 
separate evolutionary history of these two morphological components, in terms 
of when and how they acquired their present-day stellar populations, is still 
poorly understood. One reason for this is that, observationally, it is
difficult to trace the independent evolutionary history of disks and bulges, as
this requires bulge-disk decompositions to be performed on higher resolution 
images of galaxies in large statistical samples. Such analyses have been 
lacking until recently, so that studies of decomposed bulges and disks have 
been mainly restricted to small samples of highly resolved local universe 
galaxies (e.g. M\"ollenhoff et al. 1999, M\"ollenhoff \& Heidt 2001,
M\"ollenhoff 2004, Fisher \& Drory 2008, Fabricius et al. 2012). The situation
is now rapidly changing, with the advent of deep wide-field spectroscopic and
photometric surveys of galaxies (e.g. SDSS, York et al. 2000; GAMA, Driver et
al. 2011), which are providing us with large samples of galaxies for which
major morphological components can be resolved out to z=0.1. This trend will
continue into the future with the advent of new ground based surveys such as 
The VST Atlas, The Kilo Degree Survey (KiDS; de Jong et al. 2012), the Dark
Energy Survey (DES; The DES collaboration 2005), which will provide wide-field
imaging surveys with sub-arcsec resolution, and will culminate in the
wide-field diffraction-limited space-borne surveys made with EUCLID (Laureijs
et al. 2010). In parallel, automatic routines such as GALFIT 
(Peng et al. 2002, Peng et al. 2010), GIM2D (Simard et al. 2002), BUDDA
(Gadotti 2008) or MegaMorph (H\"au\ss{}ler et al. 2013, Vika et al. 2013) have 
been developed to
address the need to fit large numbers of images of galaxies with 1D analytic
functions for the characterisation of the surface brightness distributions of
their stellar components. In particular these routines allow bulge-disk
decomposition to be performed routinely, as already done by \cite{All06},
\cite{Ben07}, \cite{Cam09}, \cite{Gad09}, \cite{Sim11}, \cite{Lac12},
\cite{Bru12}, and \cite{Ber12}.

One potential problem with the results coming from bulge-disk decomposition is
that the available routines that are commonly used to perform
surface-brightness photometry cannot take into account the effects of dust. It 
is already known that spiral galaxies contain large amounts of dust (Stickel et
al. 2000, Tuffs et al. 2002, Popescu et al. 2002, Stickel et al. 2004,
Vlahakis et al. 2005, Driver et al. 2007, Dariush et al. 2011, Rowlands et al. 2012, Bourne et al. 2012, Dale et al. 2012, Grootes et al. 2013a), and that this dust changes the appearance of disks and bulges from what would be predicted based solely on their intrinsic stellar
distributions (e.g. Tuffs et al. 2004, M\"ollenhoff et al. 2006, Gadotti et
al. 2010, Pastrav et al. 2013). Since the routines available to fit galaxy
images use simple analytic functions, most commonly S\'{e}rsic functions, the
fits to real images will be imperfect, resulting in an over- or 
under-estimation of the parameters corresponding to the projected stellar 
distributions.

In Pastrav et al. (2013) we gave a detailed account of the different
applications that require accurate knowledge of the intrinsic photometric
parameters of galaxies (i.e. corrected for the effect of dust).  Here we
add relevant applications that have been emphasised by recent work. Thus,
Casaponsa et al. (2013) showed that cosmic size magnification can be used to
complement cosmic shear in weak gravitational lensing surveys, with a view to
obtaining high-precision estimates of cosmological parameters.  Thus, not only
modification of the galaxy shape (i.e. axis ratios) - a measure of the shear -
can be used in studies of weak lensing, but, for space-based data with 0.1-0.2
arcsec resolution, the size distribution of galaxies may be an important tool
for determining cosmic size magnification. In view of this, it is extremely
important to estimate the effects of dust on the scale-lengths of galaxies.
S\'{e}rsic indices of bulges are also important, as they provide a link to their
supermassive black hole. Following on the work of Graham et al. (2001) and
Graham \& Driver (2007a), recent work by Savorgnan et al. (2013) found a clear
supermassive black-hole mass - S\'{e}rsic index relation. Thus, if accurate 
S\'{e}rsic indices can be derived (corrected for the effect of dust and for 
projection effects), then these can be used to predict black hole masses in
large samples of galaxies to derive the local black hole mass function
(e.g. Graham et al. 2007) and space density (Graham \& Driver 2007b). Dust
corrections are also important on scaling relation in galaxies in
general. Thus, Grootes et al. (2013b) have recently shown that, by applying 
dust corrections from Popescu et al. (2011) and from the present paper, the 
scatter in the scaling relation specific star-formation rate versus stellar 
mass can be reduced  from 0.58 dex to 0.37 dex.

In this paper we quantify the effect of dust on the photometric parameters of
decomposed bulges and disks of spiral galaxies. As discussed in \cite{Pas12b},
this effect can be separated from the effect of dust on disks and bulges taken
individually, as seen through a common distribution of dust. Overall, when 
performing surface-brightness photometry there are three corrections that 
should be taken into account: projection effects on disks and bulges viewed
individually; the effects of dust on disks and bulges viewed individually; and 
the projection and dust effects on the disks and bulges viewed in
combination. 

Projection effects arise even in the absence of dust, causing the fitted 
functions to imperfectly recover the structure of real disks and bulges due to 
the fact that these functions describe infinitely thin templates, in contrast to
real disks and bulges, which  have a thickness. Thus, the additional vertical 
distribution of stars superimposed on the radial distribution produces 
isophotal shapes which differ from those predicted by an infinitely thin
template. The correction for projection effects on disks and bulges seen in
isolation is needed to transform the derived photometric parameters obtained 
from fitting dustless images of disks and bulges to those characterising the 
volume stellar emissivity. Corrections for this effect have been given in 
Pastrav et al. (2013).

The second type of correction that needs to be taken into account when
performing surface-brightness photometry is due to the effects of dust on 
disks and bulges when viewed individually. Such effects arise because dust 
distorts the appearance of disks and bulges. This leads to a 
discrepancy between  the derived photometric parameters of dust-attenuated 
disks and bulges and the parameters that would be derived for
disks and bulges if they could be seen at the same inclination, but in the 
absence of dust. Corrections for this discrepancy have been given
for pure disks by \cite{Mol06} (albeit without separately
considering the projection effects), and for both disks and bulges by
Pastrav et al. (2013).

The third type of correction
relates to the joint projection and dust effect on disks and bulges viewed in 
combination, attention to which was first drawn by \cite{Gad10}. This effect
causes the decomposed attenuated disk and decomposed attenuated bulge fitted
with infinitely thin and dustless templates to differ from the appearance of 
the real dust-attenuated disk and bulge.
In other words the decomposed dust-attenuated disk in the presence of a bulge 
may be imperfectly subtracted and therefore differ from the dust-attenuated 
disk that would be fitted if the galaxy were to have no bulge. Conversely, the 
decomposed dust-attenuated bulge in the presence of a disk may also be 
imperfectly subtracted and differ from how it would appear in reality 
if it could be seen in the absence of the stellar disk. Of course these
artifacts are specific to routines that perform bulge-disk decomposition using 
simple analytical infinitely thin dustless templates. However, this is the 
common practice, as it is the only feasible approach at present.

As mentioned before, in \cite{Pas12b} we quantified the projection effects and 
the effects of dust on disks and bulges viewed in isolation. Here we consider 
the joint projection and dust effects on disks and bulges viewed in 
combination, thus completing the tool kit needed to fully correct the derived 
photometric parameters. As proposed in \cite{Pas12b}, these three
effects can be multiplied (or added) together (depending on the photometric
parameter considered) using a chain correction approach, to derive total
corrections needed to convert the photometric parameters obtained from 
bulge-disk decomposition of spiral galaxies into those of the volume stellar 
emissivity. Here and in Pastrav et al. (2013) we used simulations based on a 
radiation transfer model that
can simultaneously account for both dust-attenuation in the 
ultraviolet (UV)/optical range and dust emission in the mid-infrared
(MIR)/far-infrared (FIR)/sub-millimeter (sub-mm) range. Most of the simulations
come from the library of \cite{Pop11}, while additional simulations have been
created for analysis in \cite{Pas12b}. In this paper we provide a comprehensive
data set of corrections for decomposed disks and bulges that cover the whole
parameter range in dust opacity, inclination and wavelength. The corrections
are also provided for two different values of bulge-to-disk ratios. All the 
corrections are made publically available at the CDS database.

This paper is organised as follows. In Sect.~2, we briefly describe the stellar emissivity and dust distributions used in the simulations. The method and
general approach used to fit the simulated images of the galaxies and derive
the apparent photometric parameters of the decomposed disks and bulges is
explained in Sect.~3, while the technical details of the whole fitting process
are presented in Sect.~4. In Sect.~\ref{sec:proj} we quantify the projection
effects on the bulge-disk decomposition process, while the 
dust effects on disks and bulges seen in combination are given in 
Sect.~\ref{sec:dust}. Single S\'{e}rsic fits to the same simulated images are 
presented in Sect.~\ref{sec:single}. In Sect.~\ref{sec:application} we present
an application of our predictions for the inclination dependence of dust
effects, while in Sect.~\ref{sec:summary} we summarize our results.

\section{Simulated images}

Our simulated images are those used to generate the library of 
UV/optical dust attenuations first presented in Tuffs et al. (2004) and then
in updated form in \cite{Pop11}. This library is self-consistently calculated
with the corresponding  library of dust- and polycyclic aromatic hydrocarbon 
(PAH)-emission spectral energy distributions (SEDs) given in \cite{Pop11}.
Additional simulations used in this work were presented in \cite{Pas12b}. The 
calculations are described at length in \cite{Pop11}. Here we only briefly mention their main characteristics.
All simulations were made using a modified version of the ray-tracing
radiative transfer code of \cite{Kyl87} and the dust model from \cite{Wei01} 
and \cite{Dra07} incorporating a mixture of silicates, graphites and PAH 
molecules. The
simulations were produced separately for old stellar disks, bulges and young
stellar disks, all seen through a common distribution of dust (in the disks). 
The geometrical
model of \cite{Pop11} consists of both a large scale distribution of diffuse
dust and stars, as well as a clumpy component physically associated with the
star-forming complexes. For the purpose of this study only the large scale
distribution of diffuse dust is considered, as it is this that affects the
large-scale distribution of UV/optical light determining the values of
parameters typically used in fitting surface-brightness distributions. 

The intrinsic distributions of volume stellar emissivity are described by 
exponential functions in both the radial and vertical direction for the disks, 
and by deprojected S\'{e}rsic functions for the bulges. The corresponding dust 
distributions are described by double (radial and vertical) exponential
functions  for the two dust disks of the model. A schematic representation of
the geometrical model can be found in Fig.~1 from \cite{Pop11}. All the
simulated images are sampled at 34.54 pc/pixel. The images of the 
individual morphological components were analysed in \cite{Pas12b} to 
quantify both the projection effects and the effects of dust on the photometric 
parameters of each component.

To quantify the projection and dust effects on bulge-disk decompositions, the 
simulated images of the old stellar disk and  bulges were summed to create 
simulated images of galaxies, for a set of values of disk inclination, 
wavelength, dust opacity, and bulge-to-disk ratios. The set of values span the 
whole parameter space of the model
of \cite{Pop11}. Thus simulations were produced for seven values of central 
face-on B-band optical depth $\tau_{B}^{f}$ (plus the dustless case), 21 values 
for the disk inclination $i$, and five wavelengths corresponding to the 
standard optical/near-infrared (NIR) bands B,V,I,J,K. We also consider two 
values of the bulge-to-disk ratio, $B/D=0.25,0.5$, where $B/D$ is the ratio of
the luminosity of a single dust attenuated bulge and disk. In other words $B/D$
is the apparent bulge-to-disk ratio (if both disk and bulges could be seen in 
isolation). As we will show in this paper the corrections for projection and 
dust effects on
bulge-disk decomposition only show a mild dependence on the bulge-to-disk
ratio, therefore there was no need to sample more finely the parameter $B/D$. For 
other
values of the $B/D$ ratio corrections for bulge-disk decomposition can be
obtained by interpolating between $B/D=(0,0.25,0.5)$ for disks and between
$B/D=(0.25,0.5,\infty)$ for bulges. We note that when total corrections are
derived to transform apparent parameters obtained from bulge-disk
decomposition into intrinsic parameters of the stellar volume emissivity, 
interpolation can be performed between four values of the bulge-to-disk ratio,
$B/D=(0,0.25,0.5,\infty)$, where corrections for $B/D=0$ and $B/D=\infty$
correspond to those of single disks and bulges, respectively.
  
We consider both 
exponential and de Vaucouleurs bulges. The values of the central face-on 
B-band dust optical depth cover a wide range, from almost dustless to 
extremely optically thick cases, 
$\tau_{B}^{f}=0.1,0.3,0.5,1.0,2.0,4.0,8.0$. Inclination was sampled according
to $\triangle cos(i)=0.05$, with $1-cos(i)\in[0,1]$, resulting in 21 values.

\section{Method}\label{sec:method}

We follow the same procedure observers apply in bulge-disk decompositions of
real images of galaxies and perform a multi-component fit of the simulated
images with two planar template shapes (commonly referred to as 
``infinitely thin disks''), one for each morphological
component. The functions used to describe these shapes are the exponential 
function:
\begin{eqnarray}\label{eq:exp}
\Sigma(r)=\Sigma_{0}~exp(-\frac{r}{r_{s}})
\end{eqnarray}
and  the variable-index S\'{e}rsic function
\begin{eqnarray}\label{eq:sersic}
\Sigma(r)=\Sigma_{0}~exp[-\kappa_{n} (\frac{r}{r_{e}})^{1/n}]
\end{eqnarray}
where $\Sigma_{0}$ is the central surface brightness, 
$r_{s}$ represents the exponential scale-length of the template, 
$r_{e}$ denotes the effective radius (enclosing
half the total flux) of the template, $n$ is the S\'{e}rsic index, and 
$\kappa_{n}$ is a constant, coupled with $n$ 
(Ciotti \& Bertin 1999, Graham \& Driver 2005).

We consider the following types of fits: i) fits combining the  
exponential function (Eq.~\ref{eq:exp})  and the variable-index S\'{e}rsic function (Eq.~\ref{eq:sersic}) for the disk and bulge component, respectively, and ii) fits combining two variable-index S\'{e}rsic functions for both the disk and the bulge.

As described in \cite{Pas12b}, our approach is to separate 
projection effects on disks and bulges seen individually, the effects of dust 
on disks and bulges viewed individually and the joint projection and dust
effects on disks and bulges viewed in combination
(see Eqs.~4-11 in Pastrav et al. 2013). For each of
these effects we present the results as corrections which can be used by 
observers separately or in combination. While the first two types of 
corrections were 
quantified and discussed in our previous work, in this paper we derive the 
third set of corrections, needed to quantify the influence of projection
effects and dust on the 
decomposition process. These are presented as ratios (for extrinsic quantities;
see Eqs.~\ref{eq:corr_exp_R}-\ref{eq:corr_sers_R_b} below) or differences 
(for intrinsic quantities; Eqs.~\ref{eq:corr_sers_n_d}- \ref{eq:corr_sers_n_b} 
below)  
between the fitted parameters obtained from bulge-disk decomposition in the 
presence of dust $R^{B/D}_{app,\,d}$, $R^{eff,\,B/D}_{app,\,d}$,
$R^{eff,\,B/D}_{app,\,b}$, $n^{sers,\,B/D}_{app,\,d}$, $n^{sers,\,B/D}_{app,\,b}$ 
(the measured parameters of the
decomposed disk/bulge), and 
the fitted parameters of the same disk/bulge if these were to be observed as 
single  components $R_{app,\,d}$, $R^{eff}_{app,\,d}$, $R^{eff}_{app,\,b}$,
$n^{sers}_{app,\,d}$,  $n^{sers}_{app,\,b}$  (already measured in Pastrav et
al. 2013), through the same
distribution of dust. Thus, the correction for the exponential scale-length of
the decomposed disk fitted with an exponential function, $corr^{B/D}(R_d)$,  is
\begin{eqnarray}\label{eq:corr_exp_R}
corr^{B/D}(R_{d}) & = & \frac{R^{B/D}_{app,\,d}}{R_{app,\,d}}\, ,
\end{eqnarray}
the corrections for the effective radii of decomposed disks and bulges fitted
with variable-index S\'{e}rsic functions,  $corr^{B/D}(R^{eff}_d)$ and $corr^{B/D}(R^{eff}_b)$, are
\begin{eqnarray}\label{eq:corr_sers_R_d}
corr^{B/D}(R^{eff}_{d}) & = & \frac{R^{eff,\,B/D}_{app,\,d}}{R^{eff}_{app,\,d}}
\end{eqnarray}
\begin{eqnarray}\label{eq:corr_sers_R_b}
corr^{B/D}(R^{eff}_{b}) & = & \frac{R^{eff,B/D}_{app,b}}{R^{eff}_{app,b}} \, ,
\end{eqnarray}
with \textit{d}=disk and \textit{b}=bulge, and the corrections
for the corresponding S\'{e}rsic index, $corr^{B/D}(n^{sers}_d)$ and $corr^{B/D}(n^{sers}_b)$, are
\begin{eqnarray}\label{eq:corr_sers_n_d}
corr^{B/D}(n^{sers}_{d})=n^{sers,\,B/D}_{app,\,d}-n^{sers}_{app,\,d}
\end{eqnarray}
\begin{eqnarray}\label{eq:corr_sers_n_b}
corr^{B/D}(n^{sers}_{b})=n^{sers,B/D}_{app,b}-n^{sers}_{app,b} \, .
\end{eqnarray}

We note here that the corrections measured by 
Eqs.~\ref{eq:corr_exp_R}-\ref{eq:corr_sers_n_b} include both a component due to
dust as well as a component due to projection effects. Unlike the corrections
measured on single components, it is not possible to only measure a dust effect
on the decomposition. However, we can measure pure projection effects on the
decomposition process, by comparing similar quantities without dust. 

Thus, the corresponding corrections due to pure projection effects are
presented as ratios (see Eqs.~\ref{eq:corr_proj_exp_R}-\ref{eq:corr_proj_sers_R_b} below) or differences 
(Eqs.~\ref{eq:corr_proj_sers_n_d}- \ref{eq:corr_proj_sers_n_b} 
below)
between the fitted parameters obtained from bulge-disk decomposition in the 
absence of dust $R^{B/D}_{i,\,d}$, $R^{eff,\,B/D}_{i,\,d}$,
$R^{eff,\,B/D}_{i,\,b}$, $n^{sers,\,B/D}_{i,\,d}$, $n^{sers,\,B/D}_{i,\,b}$ 
(the measured parameters of the
decomposed disk/bulge), and 
the fitted parameters of the same disk/bulge if these were to be observed as 
single  components $R_{i,\,d}$, $R^{eff}_{i,\,d}$, $R^{eff}_{i,\,b}$,
$n^{sers}_{i,\,d}$,  $n^{sers}_{i,\,b}$  (already measured in Pastrav et
al. 2013), again in the absence of dust. Thus, the correction for projection
effects on the 
exponential scale-length of the decomposed disk fitted with an exponential 
function, $corr^{proj,\,B/D}(R_d)$,  is
\begin{eqnarray}\label{eq:corr_proj_exp_R}
corr^{proj,\,B/D}(R_{d}) & = & \frac{R^{B/D}_{i,\,d}}{R_{i,\,d}}\, ,
\end{eqnarray}
the corrections for the effective radii of decomposed disks and bulges fitted
with variable-index S\'{e}rsic functions,  $corr^{proj,\,B/D}(R^{eff}_d)$ and 
$corr^{proj,\,B/D}(R^{eff}_b)$, are
\begin{eqnarray}\label{eq:corr_proj_sers_R_d}
corr^{proj,\,B/D}(R^{eff}_{d}) & = & \frac{R^{eff,\,B/D}_{i,\,d}}{R^{eff}_{i,\,d}}
\end{eqnarray}
\begin{eqnarray}\label{eq:corr_proj_sers_R_b}
corr^{proj,\,B/D}(R^{eff}_{b}) & = & \frac{R^{eff,\,B/D}_{i,\,b}}{R^{eff}_{i,\,b}} \, ,
\end{eqnarray}
with \textit{d}=disk and \textit{b}=bulge, and the corrections
for the corresponding S\'{e}rsic index, $corr^{proj,\,B/D}(n^{sers}_d)$ and $corr^{proj,\,B/D}(n^{sers}_b)$, are
\begin{eqnarray}\label{eq:corr_proj_sers_n_d}
corr^{proj,\,B/D}(n^{sers}_{d})=n^{sers,\,B/D}_{i,\,d}-n^{sers}_{i,\,d}
\end{eqnarray}
\begin{eqnarray}\label{eq:corr_proj_sers_n_b}
corr^{proj,\,B/D}(n^{sers}_{b})=n^{sers,\,B/D}_{i,\,b}-n^{sers}_{i,\,b} \, .
\end{eqnarray}

The corrections for projection effects can
be subtracted from the measurements that provide joint corrections, to
isolate pure dust effects, $corr^{dust,\,B/D}$ on the decomposition process. 
We can then write:
\begin{eqnarray}
corr^{dust,\,B/D} = corr^{B/D} - corr^{proj,\,B/D}
\end{eqnarray}

\begin{figure*}[tbh]
\begin{center}
\includegraphics[scale=0.66]{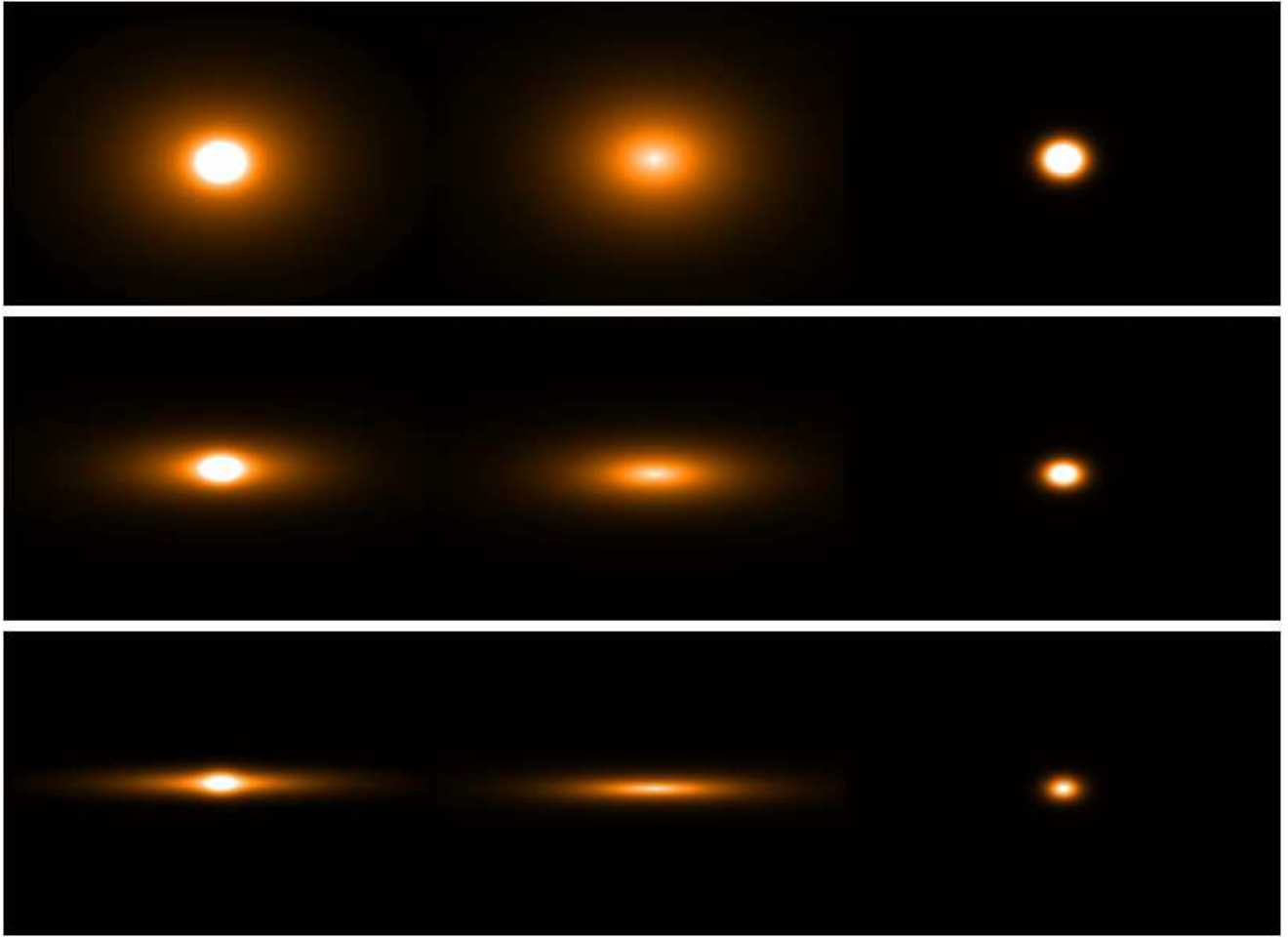} 
\caption{Simulated images of dustless galaxies with exponential bulges and
  $B/D=0.25$ (left column) and corresponding decomposed disks and bulges
  (middle and right columns). The bulge-disk decomposition fit was made with an
exponential plus a variable index S\'{e}rsic function, at inclinations 
$1-cos(i)=0.3,0.7,0.9$ ($i=46^{\circ}$ (first row), $73^{\circ}$ (second row) and 
$84^{\circ}$ (third row)).}
\label{fig:decomposed_images_exp_sers}
\end{center}
\end{figure*}

\begin{figure*}[tbh]
\begin{center}
\includegraphics[scale=1.6]{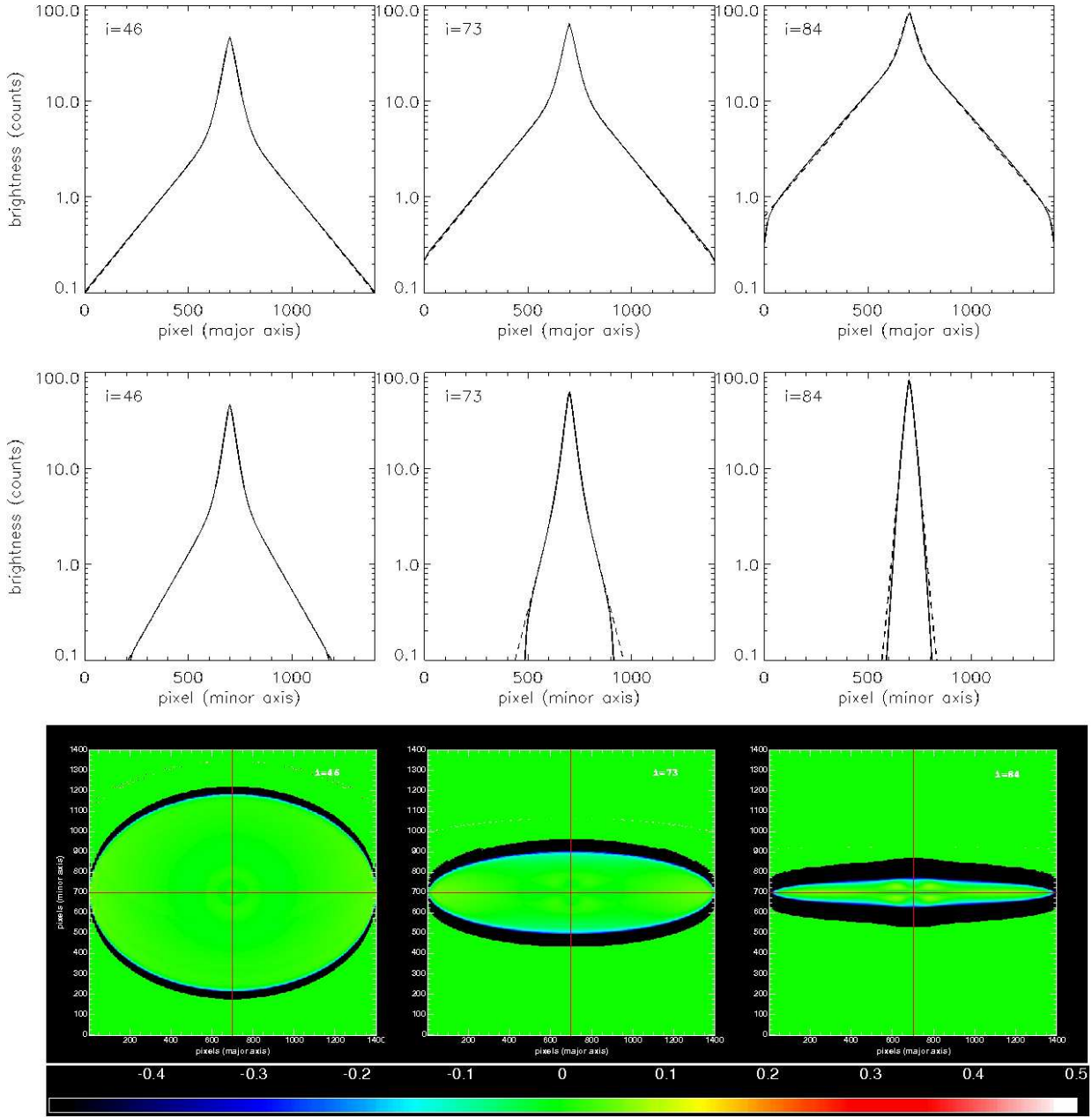} 
\caption{Major- and  minor-axis profiles of dustless galaxies (\textbf{upper 
and middle rows}) with $B/D=0.25$, in the {\bf B-band}. Fits are made with an 
{\bf exponential} function (for the {\bf disk} component) and a 
{\bf variable-index  S\'{e}rsic} function (for the {\bf exponential bulge}). 
Solid and dashed curves are for simulations and corresponding fits, 
respectively. The cuts were taken parallel and perpendicular to the major-axis 
of the simulated image, through the intensity peak, at inclinations 
$1-cos(i)=0.3,0.7,0.9$ ($i=46^{\circ},73^{\circ},84^{\circ}$).
\textbf{Lower row}: Corresponding relative residuals 
($\frac{simulation-fit}{simulation}$), at the same inclination as
the profiles. The red lines show radial and vertical cuts through the 
geometrical centre of the image.}
\label{fig:bd_profiles_resid_gal_intrin_25_b}
\end{center}
\end{figure*}
In addition to two-component fits to galaxies with two components, we 
also performed single S\'{e}rsic (sS) fits to the same simulated images 
(of galaxies with two
components). This part of our study was motivated by the fact that real images 
of galaxies are still commonly being analysed by observers using global 
S\'{e}rsic fits to obtain their radial sizes. 
Since the prime motivation for this is the derivation of disk sizes, we only 
give corrections ($corr^{sS}(R_{gal}$)) as ratios between effective radii 
obtained from single 
S\'{e}rsic fits of dusty galaxies containing bulges, and the effective radii 
of corresponding dusty disks (derived from variable-index S\'{e}rsic fits to 
the pure disks with no bulges):
\begin{eqnarray}
 corr^{sS}(R_{gal})=\frac{R_{app,\,gal}^{eff}}{R_{app,\,d}^{eff}} \, .
\end{eqnarray}
This isolates the effect of the bulge presence in constraining disk sizes from
single S\'{e}rsic fits. The correction from Eq.~14 can be used in combination with the corrections for dust and projection effects on single disks (Eq.~4 from Pastrav et al. 2013) to relate the effective radius of a disk derived from single S\'{e}rsic fits to the intrinsic effective radius of the stellar emissivity in the disk through the chain corrections:
\begin{eqnarray}
corr = corr^{proj} * corr^{dust} * corr^{sS}
\end{eqnarray}
All corrections are presented in terms of polynomial fits. The fits
are of the form:
\begin{eqnarray}\label{eq:poly}
corr(x) = \sum\limits_{k=0}^N a_k\, x^{k} & {\rm for} & 0\leq x \leq 0.95 \, ,
\end{eqnarray}
where $x=1-\cos(i)$ and N has a maximum value of 5. Although the polynomial
fits extend to 0.95, we note here that beyond $x=0.6$ the fits become
progressively poorer, due to projection effects.

\section{Fitting procedure}

Following Pastrav et al. (2013) we used the GALFIT (version 3.0.2) data analysis
algorithm (Peng et al. 2002, Peng et al. 2010) to fit our simulated images. 
GALFIT uses a non-linear least
squares fitting based on the Levenberg-Marquardt algorithm, whereby the
goodness of the fit is checked by computing the $\chi^{2}$ between the
simulated image (in the case of observations, the real galaxy image) and the
model image (created by GALFIT to fit the galaxy image). This is an iterative
process, and the free parameters corresponding to each component are adjusted
after each iteration in order to minimise the normalized (reduced) value of
$\chi^{2}$ ($\chi^{2}/N_{DOF}$, with $N_{DOF}$=number of pixels-number of free
parameters, being the number of degrees of freedom). 

Since our simulated images are noiseless, we use as input to GALFIT a 
``sigma" image (error/weight image) which is constant for all pixels, except
for points outside the physical extent of our simulated galaxies. The latter
points  were set to a very high value, to act as a mask. This was necessary 
since our simulations are truncated in their
volume stellar and dust emissivities while the fitting functions extend to 
infinity. We did not try to use the truncation functions from GALFIT, as these
truncations are for the surface-brightness distributions, rather than for the
volume stellar emissivity, as used when creating our simulated images. 
The model images have no background (by construction, unlike real images); 
accordingly the sky value was set to zero during the fitting procedure.

The free parameters of the two-component fits are: the Y coordinate of the
centre of the galaxy in pixels (while this is a free parameter, it is however
constrained to be the same for both the disk and the bulge component), the
integrated magnitudes of the disk and bulge components, the
scale-length/effective radius 
(for exponential/S\'{e}rsic function), axis-ratios, and S\'{e}rsic index (for
S\'{e}rsic function).
The axis-ratio is defined as the ratio between 
the semi-minor and semi-major axis of the model fit (for each component). The 
position angle is the angle between the semi-major axis and the Y axis 
(increasing counter clock-wise). For all our simulated images, the position angle was fixed to $-90$ (semi-major axis perpendicular on Y axis).

\section{Projection effects on the bulge-disk decomposition}
\label{sec:proj}

\begin{figure*}[tbh]
\begin{center}
\includegraphics[scale=0.385]{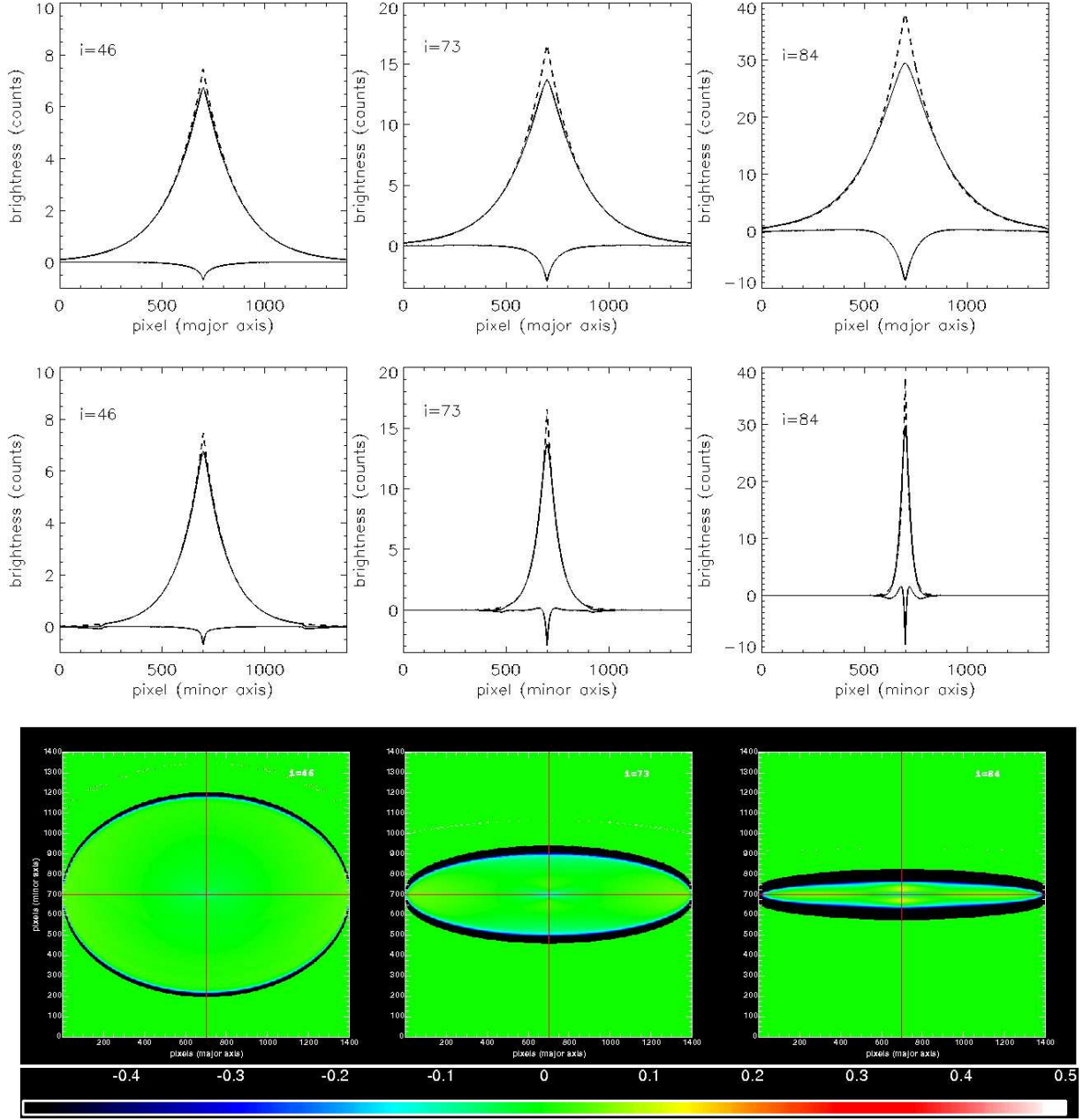} 
\caption{Major- and minor-axis profiles ({\bf upper and middle rows}) of 
simulated dust-free single disks (solid line) and of decomposed disks 
(dashed-line), for $B/D=0.25$, in the B-band. Fits are made with an 
exponential function (for the disk component) and a variable-index S\'{e}rsic
function (for the exponential bulge). The cuts were taken parallel and
perpendicular to the major-axis of the simulated image, through their
geometrical centres, at inclinations $1-cos(i)=0.3,0.7,0.9$ ($i=46,73,84$
degrees. {\bf Lower row}: Corresponding relative residuals ($\frac{simulation
  -fit}{simulation}$) at the same inclination as the profiles. The red lines show  radial and vertical cuts through the geometrical centre of the image.}
\label{fig:proj_resid}
\end{center}
\end{figure*}
As explained in Pastrav et al. (2013), even in the absence of dust, the derived 
photometric parameters of the images measured from fitting infinitely thin disk
distributions would differ from the intrinsic parameters of the volume stellar
emissivity due to the thickness (vertical stellar distribution) of real 
galaxies, which we called projection effects. These effects also act on the 
bulge-disk decomposition, causing the decomposed disks and bulges to differ in
appearance from single disks and single bulges. In other words, projection
effects are a further reason, apart from changes in morphology due to dust,
through which the decomposed
disk in the presence of a bulge may be imperfectly subtracted and therefore
differ from the disk that would be fitted if the galaxy had no
bulge. Conversely, the decomposed bulge in the presence of a disk may also be
imperfectly subtracted and differ from how it would appear in reality if it
could be seen in the absence of the stellar disk, due to projection effects. 

\subsection{Galaxies with exponential bulges}
\label{subsec:proj_expbulges}

\subsubsection{Fits with exponential + variable-index S\'{e}rsic functions}
\label{subsubsec: proj_expbulges_expsers}

The first type of fit performed on the two-component simulated dustless 
galaxies involves fitting a superposition of an exponential plus a 
variable-index S\'{e}rsic function 
for the disk and bulge component, respectively. Examples of bulge-disk 
decompositions performed in this way are given in 
Fig.~\ref{fig:decomposed_images_exp_sers}, for a bulge-to-disk ratio 
$B/D=0.25$. In Fig.~\ref{fig:bd_profiles_resid_gal_intrin_25_b} we also show 
results from these fits in the form of major- and minor-axis profiles (upper 
and middle rows) and relative residuals (bottom row).

To understand the trends due to projection effects on bulge-disk decomposition,
one needs to compare them with the similar effects produced on fits of 
single components, as described in Pastrav et al. 
(2013). Thus, we showed in our previous
paper that projection effects on single bulges manifest equally at all 
inclinations, and act to lower the measured S\'{e}rsic indices and to increase
the measured effective radii with respect to the corresponding intrinsic 
parameters of the volume stellar emissivity (see Figs.~5 and 6 from Pastrav et 
al. 2013). When a disk is also present, we will see here that the small 
deviations from the exponential form of the bulge will cause some transfer of 
light from the bulge to the disk. This will cause the
exponential fit to the disk to overpredict the amplitude of the light in the
centre and to underpredict the scale-length, and conversely will cause the fit 
to the bulge to underpredict the amplitude of the light in the centre of the
bulge and overpredict the effective radius. The effect on the integrated light
of the changes in the central amplitude outweigh the effects of the changes in
scale-length/effective radius, so that the projection effects lead to a
measured bulge-to-disk ratio slightly smaller than the one corresponding to the
bulge and the disk fitted individually.

\begin{figure}[tbh]
\begin{center}
 \includegraphics[scale=0.7]{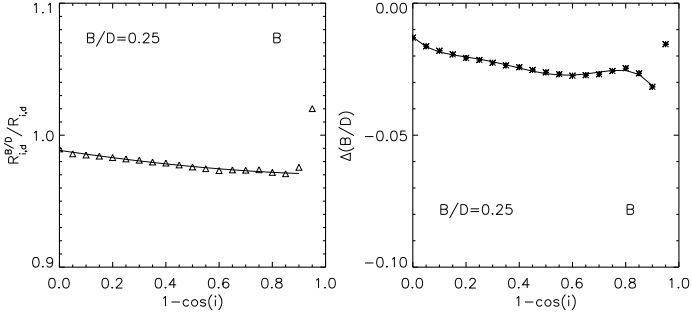}
\end{center}
\caption{\textit{Left}: Projection effects $corr^{proj,\,B/D}$ on the derived 
   scale-length of decomposed 
   {\bf disks} for $B/D=0.25$. The symbols represent the measurements while 
   the solid lines are polynomial fits to the
   measurements. The plots represent the ratio between the intrinsic 
   scale-lengths of decomposed and single disks, $R^{B/D}_{i,\,d}$ and $R_{i,\,d}$,
   respectively, as a function of inclination ($1-cos(i)$), for the B-band. 
   An {\bf exponential} (disk) {\bf plus a} {\bf variable index S\'{e}rsic} 
   (bulge) distribution were used for image decomposition.
   \textit{Right}: As in the left panel, but for the derived
  bulge-to-disk ratios, $B/D$. The effects are represented as differences 
  between the intrinsic $B/D$ of decomposed disks and bulges and those of
  single disks and bulges.}
\label{fig:proj_disk_scalelength_bd_ratios}
\end{figure}

\begin{figure}[tbh]
\begin{center}
 \includegraphics[scale=0.72]{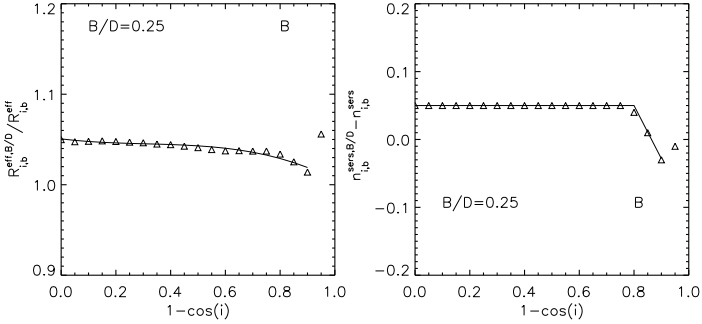}
\end{center}
\caption{As in Fig.~\ref{fig:proj_disk_scalelength_bd_ratios}, but for the
  derived effective radius $R^{eff, B/D}_{i,\,b}$  (\textit{Left}) and for the 
  derived S\'{e}rsic indices  (\textit{Right}) of decomposed {\bf exponential
  bulges}. The effects on S\'{e}rsic indices are
   represented as differences between the measured S\'{e}rsic index of 
   decomposed
   and single bulges, $n^{sers,B/D}_{i,\,b}$ and $n^{sers}_{i,\,b}$, respectively. }
\label{fig:proj_bulge_effective_radius_sersic_index}
\end{figure}

One can see the trend of overestimating the light in the centre of decomposed
disks when plotting the major- and minor-axis profiles of
the fitted decomposed disks and corresponding simulated single disks
(Fig.~\ref{fig:proj_resid}, for $B/D=0.25$). Even at low inclinations the 
fitted disks show an
excess of light in the centre (at $i=46^{\circ}$). Because the scale-length of
the fitted decomposed disk will be slightly smaller than the intrinsic
scale-length of the disk, the brightness in the outer regions of decomposed 
disks at low inclinations will be slightly underestimated, as visible in the
relative residual maps (the light yellow region in the left bottom panel of 
Fig.~\ref{fig:proj_resid}).

At higher inclinations the vertical distribution of stars starts to become
visible in the disks, producing isophotal shapes that are rounder than the
prediction of the infinitely thin disk fitting functions. In the case of 
fitting single disks, the consequence is that the fitted exponential will have
a larger scale-length than the intrinsic one (see Fig.~2 from Pastrav et al. 
2013). In the presence of a bulge, though, this trend is reversed:
additional light from the bulge is transferred to the disk, resulting in an
exponential fit with a tendentially larger amplitude and a smaller scale-length
as the disk is more inclined. As in the case of fitting single disks, the 
vertical profiles of the fitted disks will fall below the profiles of the 
simulated image over a certain range of distances from the centre,
producing the yellow wings above and below the plane in the relative residual 
maps (see right bottom panel in Fig.~\ref{fig:proj_resid}). 

To quantify these effects we compare the parameters of disks and bulges 
derived from bulge-disk decomposition of galaxies with $B/D=0.25$ with those 
obtained from fits to single 
disks and bulges. The corresponding plots showing the inclination dependence of
these projection effects are shown in
Figs.~\ref{fig:proj_disk_scalelength_bd_ratios} and 
\ref{fig:proj_bulge_effective_radius_sersic_index}. 

In Fig.~\ref{fig:proj_disk_scalelength_bd_ratios} (left) we plot the ratio of the 
scale-length of decomposed fitted disks to the scale-length of single fitted 
disks. From the definition of the plotted ratio one can immediately see that 
the trends from this figure are not directly comparable to 
those of Fig.~\ref{fig:proj_resid}, since the latter figure shows a
comparison with the simulation of a single disk rather than with the fit to a
single disk. Thus, to understand the elements of the plotted ratio one needs to
take into account both the results from Fig.~\ref{fig:proj_resid} and the
corresponding ones on the fits to single disks (Fig.~2 of Pastrav et al. 2013).
As explained above, even at low inclination the scale-length of the fitted 
decomposed disk is smaller than the intrinsic scale-length of the simulated 
single disk, and is therefore also smaller than that of the fitted single 
disks (since at low inclinations the
fitted scale-length of single disks recovers very well the intrinsic
radial scale-length of the volume stellar emissivity). Since the fitted
scale-length of the decomposed disk decreases with increasing inclination (for
low to intermediate inclinations) while the scale-length of the single disk 
increases with inclination, the ratio
of the two decreases. Towards higher inclination both the scale-length of the
fitted single and decomposed disks increase, but with the latter having a less
stronger increase, resulting in an overall ratio still decreasing with
increasing inclination. Although the scale-length is affected by
the decomposition, the derived axis-ratio of the decomposed disk is essential 
identical to that of single disks. Thus, disk axis-ratios are insensitive to
projection effects in the decomposition process (in the absence of dust).

In Fig.~\ref{fig:proj_bulge_effective_radius_sersic_index} (left) we plot the 
ratio of the effective radius of decomposed and single fitted
bulges, $R^{eff,\,B/D}_{i,\,b}$  and $R^{eff}_{i,\,b}$. As mentioned before, when
bulges are fitted in combination with a disk,
there will be a transfer of light from the bulge to the disk, resulting in a
underestimation of the light in the centre of the bulge and an overestimation 
of the effective radius. Projection effects on single bulges also manifest 
themselves in increasing the measured effective radius with respect to the 
intrinsic radius of the volume stellar emissivity. When bulge-disk 
decomposition is performed this overestimation is accentuated. Thus the ratio
$R^{eff,\,B/D}_{i,\,b}/R^{eff}_{i,\,b}$ remains supra-unity, with a small decrease
when increasing inclination. The derived S\'{e}rsic index is $\sim 0.05$ above 
the value measured on single bulges and remains constant for a large range
of inclinations (Fig.~\ref{fig:proj_bulge_effective_radius_sersic_index},
right). Since the derived S\'{e}rsic index of single bulges was found 
to be underestimated by more than $\sim 0.05$, the plot in 
Fig.~\ref{fig:proj_bulge_effective_radius_sersic_index} shows that the 
measured S\'{e}rsic index of decomposed bulges is still lower than the value 
of 1 (for the exponential bulge). 

\begin{figure*}[tbh]
\begin{center}
\includegraphics[scale=0.385]{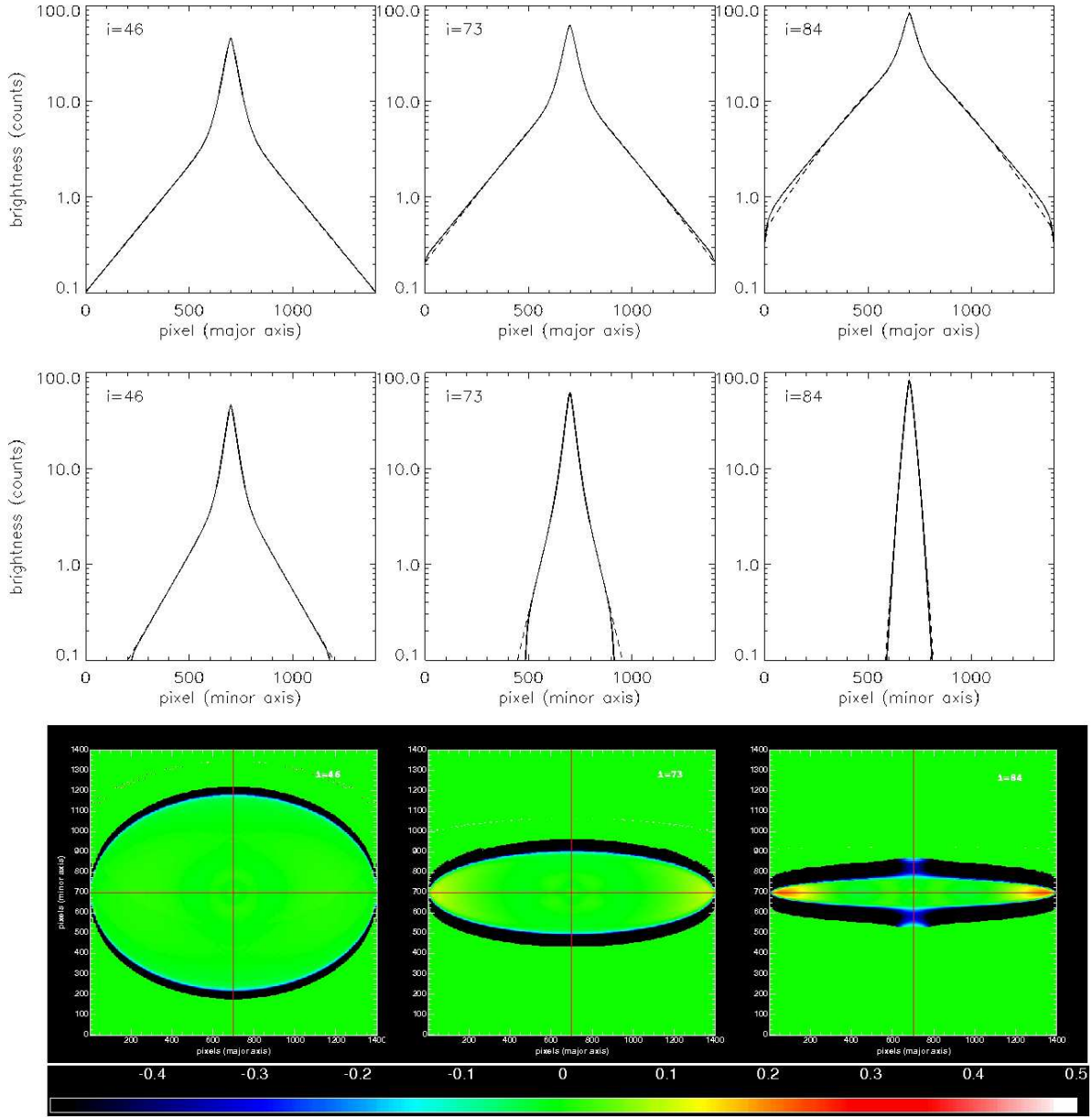} 
\caption{Major- and  minor-axis profiles of dustless galaxies (\textbf{upper 
and middle rows}) with $B/D=0.25$, in the {\bf B-band}. Fits are made with two 
{\bf variable-index  S\'{e}rsic} functions for the {\bf disk} and 
{\bf exponential bulge} components, respectively.
Solid and dashed curves are for simulations and corresponding fits, 
respectively. The cuts were taken parallel and perpendicular to the major-axis 
of the simulated image, through the intensity peak, at inclinations 
$1-cos(i)=0.3,0.7,0.9$ ($i=46^{\circ},73^{\circ},84^{\circ}$).
\textbf{Lower row}: Corresponding relative residuals 
($\frac{simulation-fit}{simulation}$), at the same inclination as
the profiles. The red lines show radial and vertical cuts through the 
geometrical centre of the image.}
\label{fig:bd_profiles_sersic_resid_gal_intrin_25_b}
\end{center}
\end{figure*}

\begin{figure*}[tbh]
\begin{center}
\includegraphics[scale=0.385]{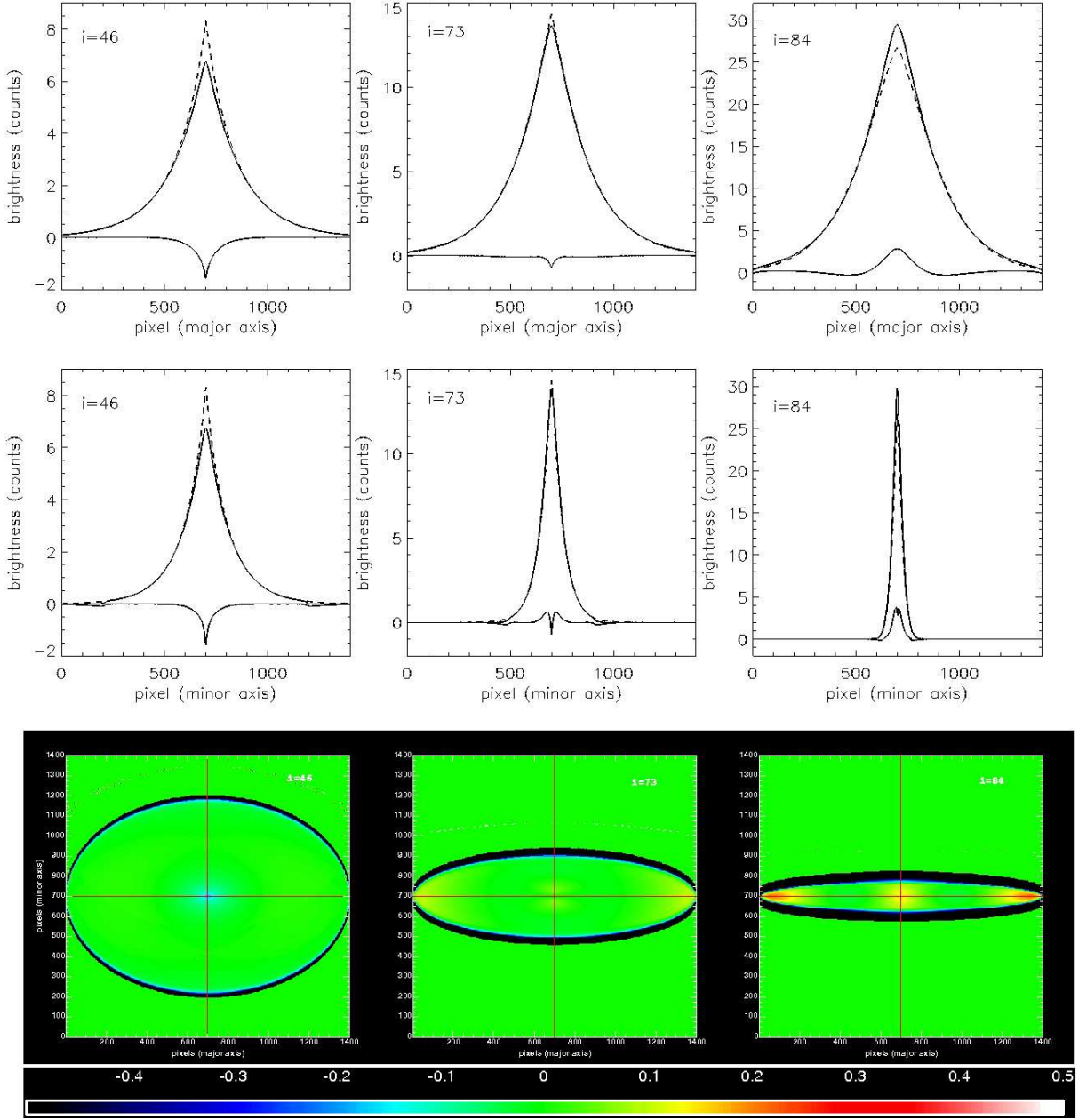}  
\caption{Major- and minor-axis profiles ({\bf upper and middle rows}) of 
simulated dust-free single disks (solid line) and of decomposed disks 
(dashed-line), for $B/D=0.25$, in the B-band. Fits are made with two 
variable-index S\'{e}rsic functions (one for disk and one for the exponential bulge component).
The cuts were taken parallel and 
perpendicular to the major-axis of the simulated image, through their geometrical
centres, at inclinations $1-cos(i)=0.3,0.7,0.9$ ($i=46,73,84$ degrees. {\bf Lower 
row}: Corresponding relative residuals ($\frac{simulation -fit}{simulation}$)
at the same inclination as the profiles. The red lines show  radial and 
vertical cuts through the geometrical centre of the image.}
\label{fig:proj_sersic_resid}
\end{center}
\end{figure*}

As expected from the trends described above, the bulge-to-disk ratio is 
slightly underestimated (Fig.~\ref{fig:proj_disk_scalelength_bd_ratios}, 
right), with the ratio showing a small decrease with increasing inclination.

The projection effects derived for fits with an exponential plus a 
variable-index S\'{e}rsic function are relatively insensitive to the value of 
 $B/D$. Thus, for an increase of the bulge-to-disk ratio to $B/D=0.5$ there is
only a $1\%$ increase in the amplitude of the correction for the
scale-length of the exponential disks. Bulges seem to be even less affected
(less than $1\%$ change in the correction), while the overall trends with
inclination remain unchanged. Examples of plots showing projection 
effects on decomposed 
disks and bulges for galaxies with $B/D=0.5$ are given in 
Appendix~\ref{sec:app_proj_exp}. 

\subsubsection{Fits with two variable-index S\'{e}rsic functions}
\label{subsubsec: proj_expbulges_serssers}

The second type of fit performed on the two-component simulated dustless 
galaxies involves fitting a superposition of two variable-index S\'{e}rsic 
functions for the disk and bulge component, respectively. In 
Fig.~\ref{fig:bd_profiles_sersic_resid_gal_intrin_25_b} we show 
results from these fits in the form of major- and minor-axis profiles (upper 
and middle rows) and relative residuals (bottom row) of galaxies with
$B/D=0.25$. Comparing these residual
maps with those obtained when fitting an exponential plus a variable-index 
S\'{e}rsic function to galaxies having the same bulge-to-disk ratio $B/D=0.25$
(Fig.~\ref{fig:bd_profiles_resid_gal_intrin_25_b}), one can see an
overall improvement in the fits at all inclinations. In particular there is an
 increased area of green colour ($\sim 0\%$ residuals). The improvement 
in the reduced $\chi^2$ is $12\%$ at $46^{\circ}$, $25\%$ at $73^{\circ}$ and 
$87\%$ at $84^{\circ}$. This is to be expected, due to the additional free 
parameter of the fit (the S\'{e}rsic index of the S\'{e}rsic function used to 
fit the disk). 

Although the overall fit is improved, the decomposed components are less
accurately extracted, due to the less constrained fit. This can be
seen in Fig.~\ref{fig:proj_sersic_resid}, where we only show residual maps 
between the decomposed fitted disks and the simulated single disks
(corresponding to Fig.~\ref{fig:bd_profiles_sersic_resid_gal_intrin_25_b}) at 
different inclinations
together with the corresponding radial and vertical profiles. In particular one
can see that at low inclinations the overestimation of the amplitude in the
centre of the disk is accentuated as compared to the situation of an
exponential fit to the disk, indicating an even more pronounced transfer of
light from the bulge to the disk. The relatively large blue region in the 
centre of the residual image for $i=46^{\circ}$ 
(left bottom panel of Fig.~\ref{fig:proj_sersic_resid})
represents an overestimation of the surface brightness of around $15\%$, 
while the corresponding panel of Fig.~\ref{fig:proj_resid} 
only shows a small blue region, with an overestimation of around 
$5-10\%$. Similar conclusions can be drawn from the corresponding profiles, 
which also show that the brightness of the decomposed disk (plotted as dotted 
line) exceeds that of the simulated single disk (solid line), in the centre.

At higher inclinations, the yellow wings that were seen above and below the 
plane in the residual maps of disks decomposed with exponential functions 
(right bottom panel of Fig.~\ref{fig:bd_profiles_resid_gal_intrin_25_b}) 
now merge into a
region of continuous yellow colour in the centre of the disk, as seen in the 
right bottom panel of Fig.~\ref{fig:proj_sersic_resid}. This means that at 
high inclinations the
surface-brightness in the centre regions of decomposed disks fitted with 
S\'{e}rsic functions will be underestimated by around $15\%$. Thus, at higher
inclinations there is a transfer of light from the disk to the bulge (see also
right top and middle panels of Fig.~\ref{fig:proj_sersic_resid}). In 
addition the outer regions of highly inclined decomposed disks fitted with 
S\'{e}rsic functions is less well fitted in comparison with decomposed disks 
fitted with exponential functions.

\begin{figure}[tbh]
\begin{center}
 \includegraphics[scale=0.72]{proj_eff_sersic_sersic_eff_radius_disk_sersic_index_intrin_25_b.epsi}
\end{center}
\caption{\textit{Left}: Projection effects $corr^{proj,\,B/D}$ on the derived 
   effective radius of 
   decomposed {\bf disks} for $B/D=0.25$. The symbols represent the 
   measurements while the solid lines are polynomial fits to the
   measurements. The plots represent the ratio between the intrinsic 
   effective radius of decomposed and single disks, $R^{eff,\,B/D}_{i,\,d}$ and 
   $R^{eff}_{i,\,d}$, respectively, as a function of inclination ($1-cos(i)$),
   for the B-band. 
   Two variable index S\'{e}rsic functions were used for image decomposition.
   \textit{Right}: As in the left panel, but for the derived
  S\'{e}rsic index, $n^{sers}$. The effects are represented as differences 
  between the measured S\'ersic index of decomposed and single disks, 
  $n^{sers,\,B/D}_{i,\,d}$ and $n^{sers}_{i,\,d}$, respectively.}
\label{fig:proj_disk_effective_radius_sersic_index}
\end{figure}

\begin{figure}[tbh]
\begin{center}
 \includegraphics[scale=0.72]{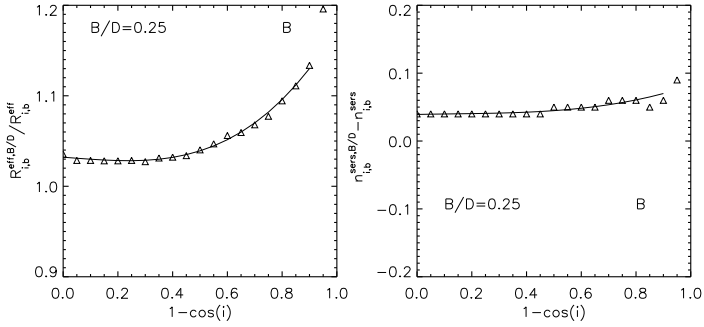}
\end{center}
\caption{As in Fig.~\ref{fig:proj_disk_effective_radius_sersic_index}, but for
 the decomposed {\bf exponential bulges}.}
\label{fig:proj_bulge_sers_sers_effective_radius_sersic_index}
\end{figure}

\begin{figure}[tbh]
\begin{center}
 \includegraphics[scale=0.36]{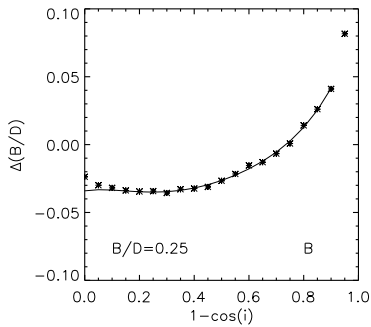}
\caption{Projection effects $corr^{proj,\,B/D}$ on the derived bulge-to-disk
  ratios, $B/D$. The effects are represented as differences 
  between the intrinsic $B/D$ of decomposed disks and bulges and those of
  single disks and bulges.
  Fits are done with {\bf two variable S\'{e}rsic index functions}.}
\label{fig:sersic_sersic_bd_ratios_difference_galaxy_intrin_exp_bulge_25_b}
\end{center}
\end{figure}

To derive the projection effects on the parameters of disks and bulges
decomposed from fitting two variable-index S\'{e}rsic functions, we compare again the results
of the fits with those obtained for single disks and bulges (individually 
fitted with variable-index S\'{e}rsic functions in Pastrav et al. 2013). In 
Fig.~\ref{fig:proj_disk_effective_radius_sersic_index} we show the projection
effects of decomposed disks, for galaxies with $B/D=0.25$. Close to face-on 
inclinations the derived
effective radius of the decomposed disk is slightly smaller than that derived
for single disks, the latter being a good match to the intrinsic effective
radius of the volume stellar emissivity. This results in a ratio 
$R^{eff,\,B/D}_{i,\,d}/R^{eff}_{i,\,d}$ which is slightly less than one at
$i=0^{\circ}$ (see Fig.~\ref{fig:proj_disk_effective_radius_sersic_index}
left), similar to the results obtained when performing bulge-disk
decomposition with exponential plus S\'{e}rsic functions (see 
Fig.~\ref{fig:proj_disk_scalelength_bd_ratios}, left). With increasing
inclination the effective radius of decomposed disks increases, following the 
transition between an overestimation of the central 
surface-brightness (light transfer from bulge to disk) to an underestimation 
of the central surface-brightness (light transfer from disk to bulge). Since
the effective radius of single disks fitted with variable S\'{e}rsic index
functions decreases with increasing inclination (see Fig.~2
from Pastrav et al. 2013), the overall trend of the ratio 
$R^{eff,\,B/D}_{i,\,d}/R^{eff}_{i,\,d}$ is to increase with increasing inclination.

The slight underestimate in the effective radius of the decomposed disk at low
inclinations is also accompanied by a slight  overestimate of the derived 
S\'{e}rsic index (which takes the value of $\sim 1.07$). Since the fitted 
S\'{e}rsic index of a single 
disk seen at low inclinations exactly matches the value of 1 (corresponding to 
an exponential disk), the projection effects manifest in a positive correction
for the S\'{e}rsic index measured for face-on disks (see right panel of 
Fig.~\ref{fig:proj_disk_effective_radius_sersic_index}). With increasing inclination,
both the derived S\'{e}rsic index of single disks and of decomposed disks
decreases, but at different rates, such that a decreasing trend in the 
correction $n^{sers,\,B/D}_{i,\,d}-n^{sers}_{i,\,d}$ is produced.

The projection effects on decomposed bulges are shown in
Fig.~\ref{fig:proj_bulge_sers_sers_effective_radius_sersic_index}. The
effective radius of the decomposed bulge is always larger than that of a 
single bulge, with the ratio of the two increasing with increasing 
inclination. The derived S\'{e}rsic index is slightly larger than that of a 
single bulge, and remains essentially constant with increasing inclination.

The bulge-to-disk ratio of decomposed disks and bulges at low inclinations is 
slightly smaller than the one derived for single components (see
Fig.~\ref{fig:sersic_sersic_bd_ratios_difference_galaxy_intrin_exp_bulge_25_b}).
This is to be expected, since, as explained above, light from the bulge is 
transferred to the disk in the fitting process. This behaviour is similar to the
one encountered when doing fits with an exponential plus a S\'{e}rsic function,
since in both cases the surface-brightness distribution in the central regions
of disks is overestimated. At higher inclinations however, an opposite trend
is observed, with a bulge-to-disk ratio of decomposed disks and bulges slightly
larger than the one derived from single components. Essentially the correction
for projection effects on $B/D$ increases smoothly with inclination. 
This behaviour is different from the trend obtained in the case of an 
exponential plus a S\'{e}rsic index fit. 

The projection effects derived for fits with two variable-index S\'{e}rsic 
function are also relatively insensitive to the value of $B/D$. Thus, for 
an increase of the bulge-to-disk ratio to $B/D=0.5$ there will only be a $1\%$ 
increase in the amplitude of the corrections for the disks. Examples of plots 
showing projection effects on decomposed disks and bulges for galaxies with 
$B/D=0.5$ are given in the Appendix~\ref{sec:app_proj_exp}.

\subsection{Galaxies with de Vaucouleurs bulges}

When modelling galaxies containing higher S\'{e}rsic index bulges, including 
de Vaucouleurs bulges, one of the main factors shaping projection 
effects is the truncation radius of the bulge. Unfortunately this parameter is
unknown from observations. From simulations we find that the radial stellar 
profiles at large galactocentric radii starts to be dominated by the light from
the bulge instead of the disk, if galaxies contain bulges with un-truncated
stellar distributions. For a galaxy with a de Vaucouleurs bulge, a truncation of
the bulge at 3 effective radii is enough to circumvent this problem. For
galaxies with higher than $n^{sers}=4$ the truncation of the bulge would 
need to be at less than 3 effective radii. Overall the truncation radius would
depend in this case on the S\'{e}rsic index of the bulge. Whether this has any
bearing to reality it is unknown. Overall this pins down to the underlying
problem that we do not know what the intrinsic distribution of the volume
emissivity of the bulge is, and that there is no physical interpretation
attached to the S\'{e}rsic distribution that is used to described the projected
stellar distribution (images) of bulges. The deprojected S\'{e}rsic distribution
does not have an exact analytic formula due to the singularity in the centre,
and therefore approximate formulae have been proposed to describe the volume
stellar emissivity. In our model we consider an analytic formula that, when
integrated to infinity reproduces the S\'{e}rsic distribution of a 2D
map. Nonetheless, if bulges are truncated, and we insist in preserving the same
analytic formulation, we end up in simulations that are not perfectly fitted by
S\'ersic distributions.\footnote{As shown in Pastrav et al (2013), the shorter 
the truncation radius is, the larger the deviation from the S\'{e}rsic
distribution.} We included these deviations in our projection effects, although,
unlike the case of the disk, this is a reverse problem to the disk: in disks we
know the intrinsic volume emissivity and therefore we can predict and compare
with observations the projected distribution, while in bulges we know the 
projected stellar emissivity, but we cannot exactly predict the intrinsic
stellar emissivity, and therefore we cannot directly compare with
observations. 
Unlike the disk, it is therefore unclear whether what we call projection
effects on bulges is a real effect or just a
limitation of our knowledge of the true 3D stellar distribution of
bulges. In composite systems with disk and bulges, the combined projection
effects of disk and bulges that act on the bulge-disk decomposition amplifies
if the truncation radius of the bulge is short, as is the case for de
Vaucouleurs bulges in our model. This produces projection effects that are 
larger than in the case of exponential bulges, which we truncated at 10 
effective radii. We
caution therefore the reader that corrections for projection effects derived
for systems with de Vaucouleurs bulges are less certain. 

\begin{figure*}[tbh]
\hspace{0.75cm}
\includegraphics[scale=0.48]{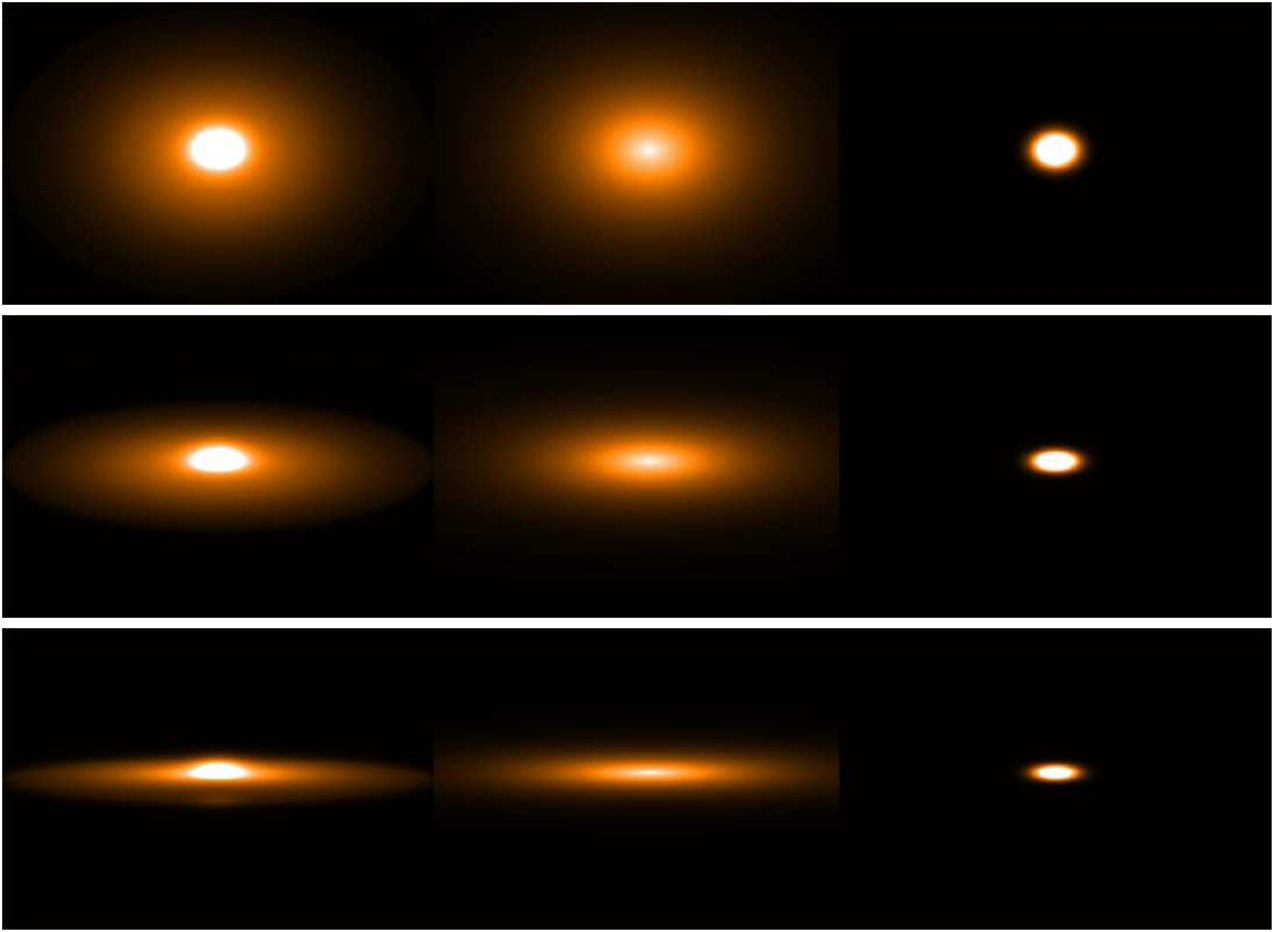}  
\caption{\label{fig:decomposed_exp_sersic_bulge_disk_images_25_b_high_op}
  Simulated images of galaxies with {\bf exponential bulges} and 
$B/D=0.25$ (left column) and 
corresponding decomposed disks and bulges (middle and right columns). The 
bulge-disk decomposition fit was made with an {\bf exponential plus a variable index
S\'{e}rsic} function, at inclinations $1-cos(i)=0.3,0.7,0.9$ 
($i=46^{\circ}$ (first row), $73^{\circ}$ (second row) and 
$84^{\circ}$ (third row)), for $\tau_{B}^{f}=4.0$.} 
\end{figure*}

\begin{figure*}[tbh]
\begin{center}
\includegraphics[scale=0.385]{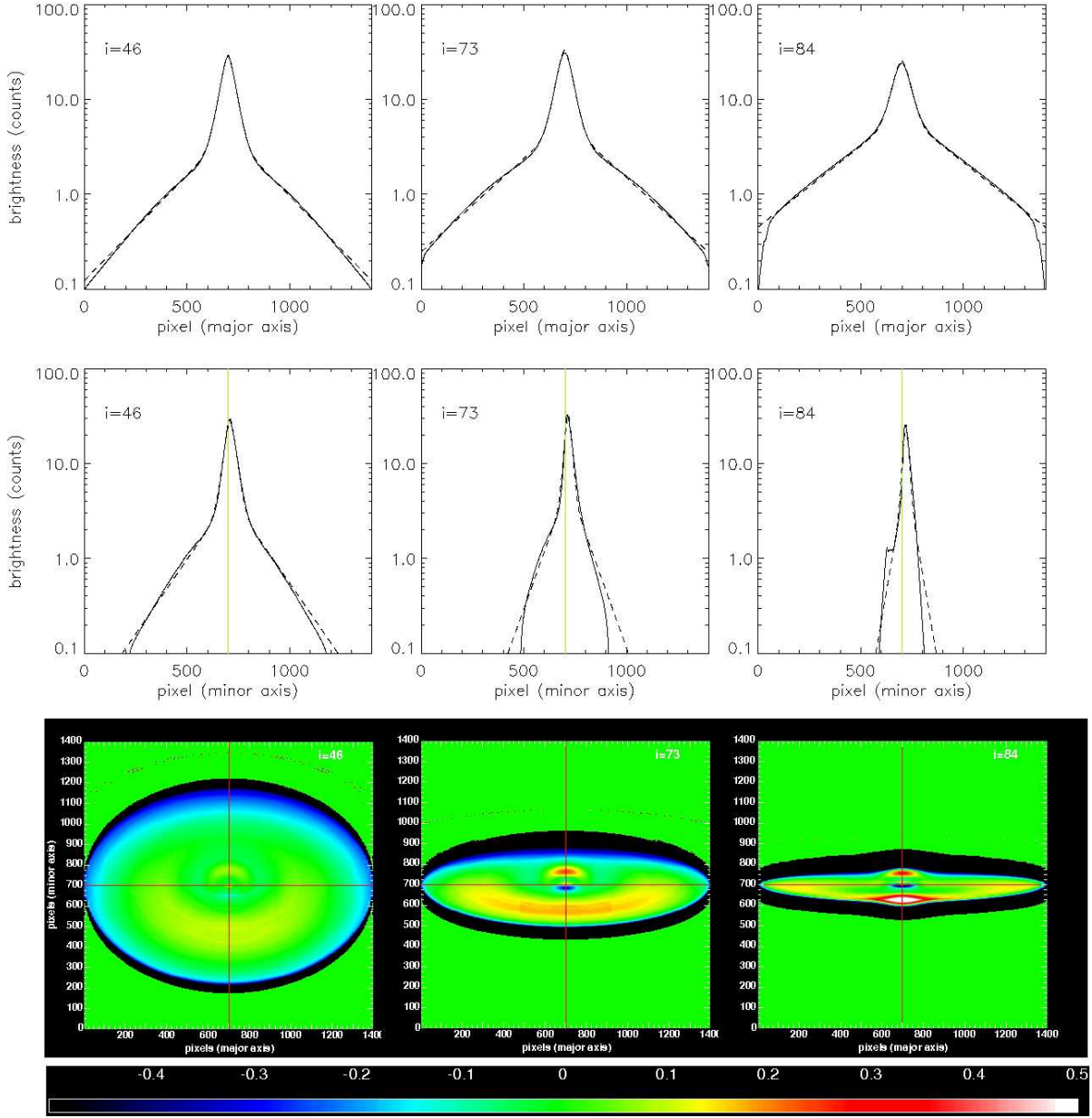}  
\caption{\label{fig:bd_profiles_relative_residuals_obs_25_b_high_op} Major- and
  minor-axis profiles of dusty galaxies (\textbf{upper and middle rows}) with 
$B/D=0.25$, and $\tau_{B}^{f}=4.0$, in the {\bf B-band}. Fits are made with an 
{\bf exponential} function (for the {\bf disk} component) and a {\bf variable-index
  S\'{e}rsic}  
function (for the {\bf exponential bulge}). Solid and dashed curves are 
for simulations and corresponding fits, respectively. The cuts were taken 
parallel and perpendicular to the major-axis of the simulated image, through 
the intensity peak, at inclinations $1-cos(i)=0.3,0.7,0.9$ 
($i=46^{\circ},73^{\circ},84^{\circ}$). The light green line 
shows a cut through the geometrical centre of the image.
\textbf{Lower row}: Corresponding relative residuals 
($\frac{simulation-fit}{simulation}$), at the same inclination and opacity as
the profiles. The red lines show radial and vertical cuts through the 
geometrical centre of the image.}
\end{center}
\end{figure*}

\begin{figure*}[tbh]
\begin{center}
\includegraphics[scale=0.385]{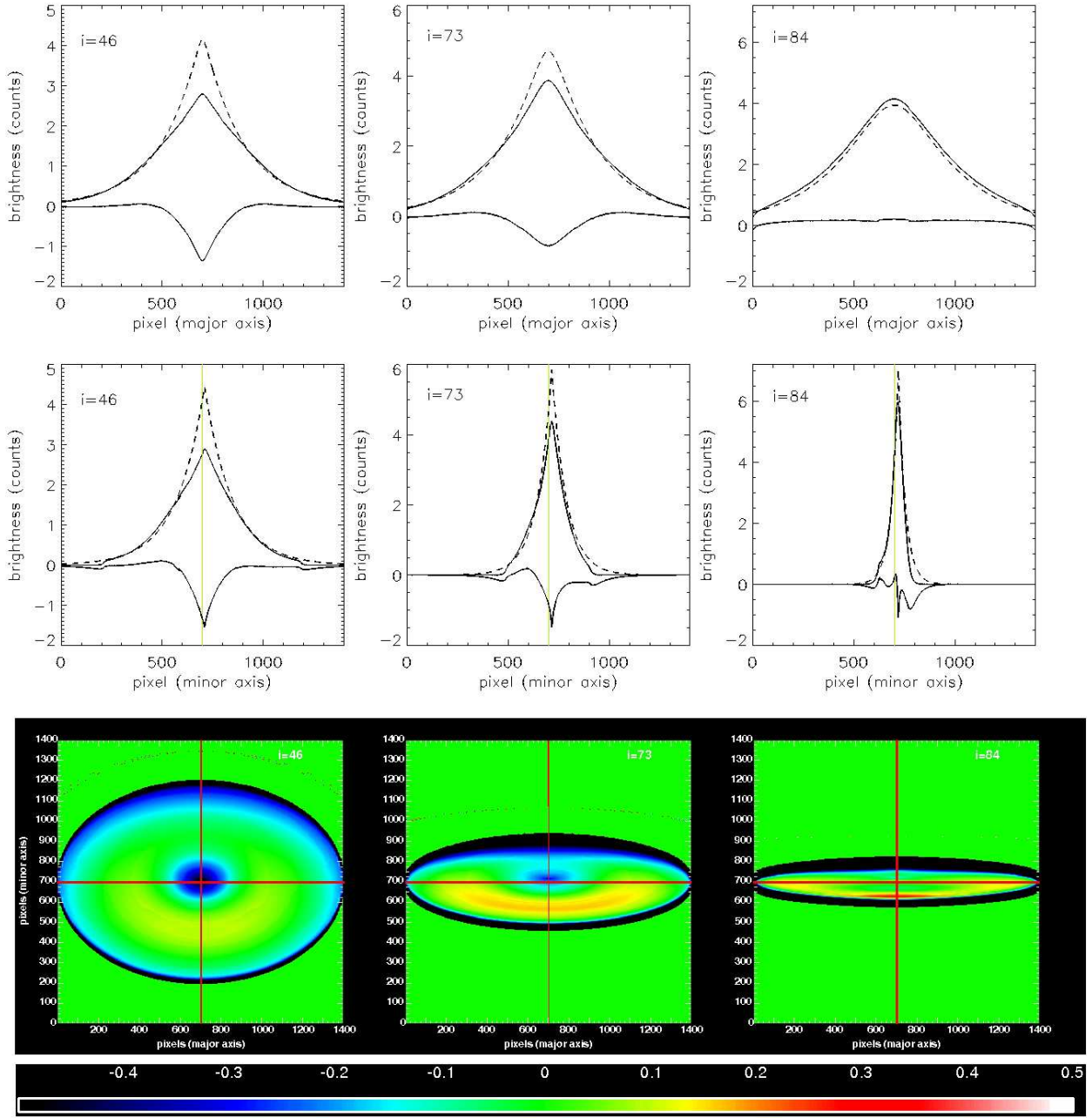}  
\caption{\label{fig:bd_disk_decomp_profiles_residuals_obs_25_b} 
 Major- and minor-axis profiles ({\bf upper and middle rows}) of 
simulated dusty single disks (solid line) and of decomposed dusty disks 
(dashed-line), for $B/D=0.25$ and $\tau_{B}^{f}=4.0$, in the B-band. Fits are made with an 
exponential function (for the disk component) and a variable-index S\'{e}rsic
function (for the exponential bulge). The cuts were taken parallel and 
perpendicular to the major-axis of the simulated image, through their geometrical
centres, at inclinations $1-cos(i)=0.3,0.7,0.9$ ($i=46,73,84$ degrees.
The light green line  shows a cut through the geometrical centre of the image. 
{\bf Lower row}: Corresponding relative residuals ($\frac{simulation -fit}{simulation}$),
at the same inclination and opacity as the profiles. The red lines show 
radial and vertical cuts through the geometrical centre of the image.}
\end{center}
\end{figure*}

Following the same procedure as in Sect.~\ref{subsec:proj_expbulges}, we derive
corrections for projection effects both for decompositions involving an
exponential plus a variable index S\'{e}rsic function and fits with two variable
index S\'{e}rsic functions. Examples of corresponding plots with 
corrections are given in Appendix~\ref{sec:app_proj_dev}.

\section{The effects of dust on the bulge-disk decomposition}
\label{sec:dust}

In this section we present and discuss the effects of dust on the process of 
decomposing galaxy images and therefore on the photometric parameters of
decomposed disks and bulges. As mentioned before, unlike the
corrections measured on single components, it is not possible to only
measure a dust effect on the decomposition. The measurements are for
the joint effect of dust and projections on the
decomposition. Thus, using Eqs.~\ref{eq:corr_exp_R}-\ref{eq:corr_sers_n_b}, 
we relate the measured photometric parameters of decomposed disks and
bulges to those obtained in our previous study (the apparent values from fitting 
individual components) in order to quantify $corr^{B/D}$, the dust
and projections effects on 
the bulge-disk decomposition. Then, by subtracting the corrections for
projection effects (as described in Sect. ~\ref{sec:proj}), we can isolate the
pure dust effects, $corr^{dust,\,B/D}$. These effects are quantified for galaxies 
with exponential bulges (Sect.~\ref{subsec:expbulges}) and de Vaucouleurs bulges
(Sect.~\ref{subsec:devaucbulges}). 

One of the main problems when performing bulge-disk decomposition of dusty 
galaxies is the dust-induced asymmetries in the surface-brightness 
distributions, in particular at higher inclinations. These asymmetries are 
present in 
both the dust-attenuated disk and bulge, as described in Pastrav et
al. (2013), and because of them, the position of the intensity peak does not
coincide with the geometrical centre of the image. In addition, 
the position of the peak intensity of each dust-attenuated component is 
differently shifted from the geometrical centre. Therefore the combined image
will have a peak intensity which will coincide neither with the geometrical
centre, nor with the true position of the peak intensity of either disk or 
bulge. As a consequence, the resulting 
bulges and disks will be imperfectly subtracted when performing bulge-disk 
decomposition with simple analytic templates, irrespective of the 
combination of functions used to fit the composite systems 
(exponential plus S\'{e}rsic or S\'{e}rsic plus S\'{e}rsic). 

\subsection{Galaxies with exponential bulges}\label{subsec:expbulges}
\subsubsection{Fits with exponential + variable-index S\'{e}rsic functions}
\label{subsubsec: expbulges_expsers}

The first type of fit performed on the two-component simulated galaxies 
consists of an exponential plus a variable-index S\'{e}rsic function for the
disk and bulge component, respectively. Examples of bulge-disk decompositions
performed in this way are given in 
Fig.~\ref{fig:decomposed_exp_sersic_bulge_disk_images_25_b_high_op}. 
In Fig.~\ref{fig:bd_profiles_relative_residuals_obs_25_b_high_op} we also show 
results from these fits in the form of
major- and minor-axis profiles (upper and middle rows) and relative residuals
(bottom row). One can see 
the afore-mentioned asymmetries about the major-axis, which 
increase with increasing inclination of the disk. The
blue region of negative residuals in the outer disk seen in
Fig.~\ref{fig:bd_profiles_relative_residuals_obs_25_b_high_op} is due to the 
fact that the simulations are
truncated while the fits extend to infinity.

\begin{figure}[tbh]
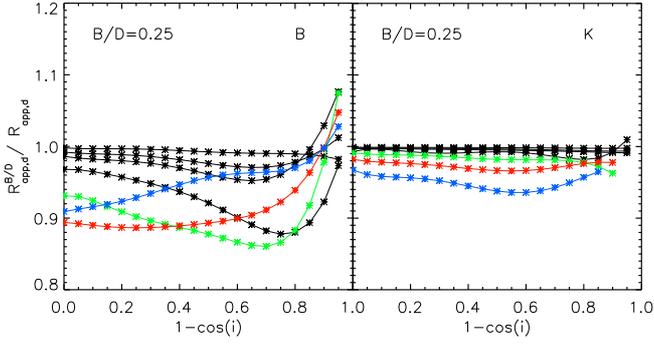

 \includegraphics[scale=0.38]{proj_corr_exp_sersic_bulge_disk_decomp_scale_length_vs_inclination_obs_25_b.epsi}
 \hspace{-0.33cm}
 \includegraphics[scale=0.38]{proj_corr_exp_sersic_bulge_disk_decomp_scale_length_vs_inclination_obs_25_k.epsi}
 \caption{\label{fig:bd_exp_sersic_scale_lengths_obs_25} Dust effects
   ($corr^{dust,\,B/D}$) on the derived scale-length of decomposed {\bf disks} for 
   $B/D=0.25$. The solid lines are polynomial fits to the measurements.
   The plots represent the ratio between the apparent 
   scale-lengths of decomposed and single disks, $R^{B/D}_{app,d}$ and $R_{app,d}$,
   respectively (corrected for projection effects), as a function of 
   inclination ($1-cos(i)$), for the B and K 
   optical bands. An {\bf exponential} (disk) {\bf plus a} {\bf variable index 
   S\'{e}rsic} (bulge) distribution were used for image decomposition. 
   The black curves are plotted for $\tau_{B}^{f}= 0.1,0.3,0.5,1.0$,
   while the other curves correspond to $\tau_{B}^{f}=2.0$ (green), $4.0$ (red) and $8.0$ (blue).}
\end{figure}

Another effect which influences the decomposition is the flattening of the 
radial
profiles in the inner regions of dust-attenuated disks, in particular for 
higher values of dust opacity, as already discussed in Pastrav et al. (2013). 
When such disks exist
in isolation (without a bulge) and are fitted with an exponential function, the
depression of the surface-brightness in the centre of disks results in a fit
with an exponential model having a larger scale-length than the
intrinsic one. However, in the presence of a bulge, the flattening of the disk 
profile in the centre is wrongly compensated for by the fitting routine with 
stellar light from the bulge. 
This can be seen in Fig.~\ref{fig:bd_disk_decomp_profiles_residuals_obs_25_b}, where
we plot examples of relative residuals between simulated single dusty disks and
corresponding decomposed disks. The blue region in the centre (for
$i=46^{\circ}$ and $73^{\circ}$) is due to the exponential form of the
decomposed disk which rises above the flattened central region of the 
simulated attenuated single disk. 
At lower dust opacities, when the flattening of the disk is small and happens
within one effective radius of the bulge, the routine will transfer enough
light from the bulge to reasonably compensate for the flattening of the disk. 
Therefore the derived scale-length is closer (or slightly smaller) to the
intrinsic scale-length of the disk (measured at the same inclination in the
absence of dust). At higher optical depth though, when the disk is
optically thick until large radii, beyond the effective radius of the bulge,
there is still a transfer of light from the bulge to the disk, but not enough 
to compensate for the more pronounced flattening. Therefore, to
account for the remaining depression in the surface-brightness, the routine 
will tend to overestimate the scale-length of the decomposed disk (with respect
to the dustless case), as in the 
case of a single disk analysis. However, the overestimation will be smaller
than in the case of a single disk. To conclude, the derived scale-length of a
decomposed disk is close to the intrinsic one at smaller opacities and is
overestimated at higher opacities. We note here that this effect is not
visible in Fig.~\ref{fig:bd_disk_decomp_profiles_residuals_obs_25_b}, since 
the outer
regions of the disks in the residual maps are dominated by the difference
between the truncated simulation and the untruncated model. However, in all
cases the derived scale-lengths of decomposed disks will be smaller than the
derived scale-lengths in the absence of a bulge (see 
Fig.~\ref{fig:bd_exp_sersic_scale_lengths_obs_25}). 

Conversely, the decomposed 
bulge will have a slightly flatter profile in the centre than in reality, since light from the
simulated bulge
has been transferred to the simulated disk, resulting in a fit with a
smaller S\'{e}rsic index than in the case of a pure attenuated bulge (see
Fig.~\ref{fig:bd_bulge_sersic_index_difference_obs_25}). As expected, the corrections
$corr^{dust,\,B/D}$ are larger in the B-band than in the K-band. The derived 
effective radius of bulges is smaller than that of single
attenuated bulges, (see
Fig.~\ref{fig:bd_exp_sersic_eff_rad_obs_25}).

Since in the decomposition process light from the bulge is transferred to the
disk, in particular for higher opacity, the derived bulge-to-disk ratio will be
smaller than the bulge-to-disk ratio of a single bulge and a single disk
attenuated by the same dust opacity. This can be seen in
Fig.~\ref{fig:proj_corr_exp_sersic_bulge_disk_ratios_obs_25}, where the
dust correction takes negative values. 
 
The results from Figs.~\ref{fig:bd_exp_sersic_scale_lengths_obs_25} to 
\ref{fig:proj_corr_exp_sersic_bulge_disk_ratios_obs_25} are for a $B/D=0.25$. The same analysis
performed on simulations having $B/D=0.5$ show very little differences in the
results (see examples in Appendix~\ref{sec:app_dust_exp}). For disks the
amplitude of the effects slightly increases with
increasing $B/D$. For bulges the amplitude of the effects decreases
with increasing $B/D$. The trends with inclinations remain the same.

\subsubsection{Fits with two variable-index S\'{e}rsic functions}
\label{subsubsec:expbulges_serssers}

For bulge-disk decomposition performed with two variable-index S\'{e}rsic
functions there is an extra free parameter for fitting the disk
component, namely the S\'{e}rsic index of the disk. This results in an overall
better fit for the composed system. In addition, unlike the dustless case, the
decomposed disk and bulge are also better fitted in this way, with a less
transfer of light from the bulge to the disk, resulting in a solution which is
closer to the single disk and single bulge cases. This is because dust flattens
the S\'{e}rsic profiles in the centre of bulges and disks, making the 
decomposition less ambiguous and degenerate than in the dust free case, at
least for the low and intermediate inclination cases.
 As shown in Pastrav et al. (2013), the flattening of the central
parts of single disks due to attenuation is fitted with a  S\'{e}rsic index
having a lower value than the intrinsic one. When a bulge is also present,
GALFIT will find a solution with a slightly larger S\'{e}rsic index than for
the single disk (see
Fig.~\ref{fig:bd_bulge_sersic_disk_sersic_index_difference_obs_25}), because
light transfer from the bulge still occurs for all opacities. Because
of this the derived effective radii will be close (or slightly smaller) to the ones derived for single disks, as shown in Fig.\ref{fig:bd_sersic_sersic_disk_eff_rad_obs_25}.

\begin{figure}[h]
 \includegraphics[scale=0.38]{proj_corr_exp_sersic_bulge_disk_decomp_nbulge_index_variation_obs_25_b.epsi}
 \hspace{-0.33cm}
 \includegraphics[scale=0.38]{proj_corr_exp_sersic_bulge_disk_decomp_nbulge_index_variation_obs_25_k.epsi}
\caption{\label{fig:bd_bulge_sersic_index_difference_obs_25} 
As in Fig.~\ref{fig:bd_exp_sersic_scale_lengths_obs_25}, but for the derived 
S\'{e}rsic index of decomposed {\bf exponential bulges}. The effects are
represented as differences between the derived S\'{e}rsic index of decomposed and single 
bulges, $n_{app,b}^{sers,B/D}$ and  $n_{app,b}^{sers}$, respectively.} 
\end{figure}

\begin{figure}[tbh]
 \includegraphics[scale=0.38]{proj_corr_exp_sersic_bulge_disk_decomp_effective_radius_vs_inclination_obs_25_b.epsi}
 \hspace{-0.33cm}
 \includegraphics[scale=0.38]{proj_corr_exp_sersic_bulge_disk_decomp_effective_radius_vs_inclination_obs_25_k.epsi}
 \caption{\label{fig:bd_exp_sersic_eff_rad_obs_25} 
As in Fig.~\ref{fig:bd_exp_sersic_scale_lengths_obs_25}, but for the derived 
effective radius of decomposed {\bf exponential bulges}.}
\end{figure}

\begin{figure}[tbh]
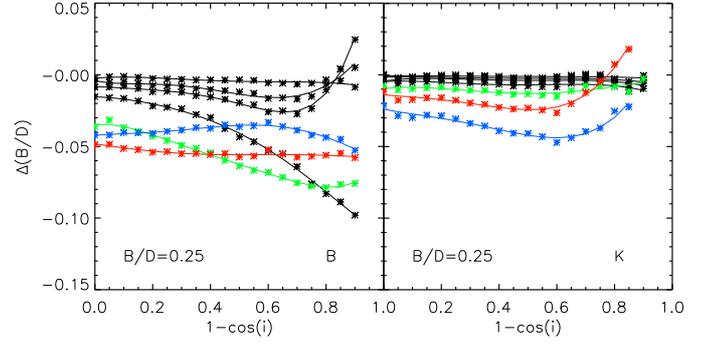

 \includegraphics[scale=0.38]{proj_corr_exp_sersic_bulge_disk_ratios_obs_25_b.epsi}
\hspace{-0.33cm}
\includegraphics[scale=0.38]{proj_corr_exp_sersic_bulge_disk_ratios_obs_25_k.epsi}
\caption{\label{fig:proj_corr_exp_sersic_bulge_disk_ratios_obs_25}
Dust effects ($corr^{dust,\,B/D}$) on the derived bulge-to-disk
ratios, $B/D$. The effects are represented as differences between the apparent
$B/D$ of decomposed disks and bulges as those of single disks and bulges. 
An {\bf exponential} (disk) {\bf plus a} {\bf variable index 
   S\'{e}rsic} (bulge) distribution were used for image decomposition. 
  The curves are plotted for $\tau_{B}^{f}= 0.1,0.3,0.5,1.0$ (from 
   the top towards the bottom), while the other curves correspond to  
   $\tau_{B}^{f}=2.0$ (green), $4.0$ (red) and $8.0$ (blue).}
\end{figure}

The effective radii (see
Fig.~\ref{fig:bd_sersic_sersic_bulge_eff_rad_obs_25}) and  the derived S\'{e}rsic index
(see Fig.~\ref{fig:bd_bulge_sersic_sersic_bulge_index_difference_obs_25}) of the bulge is relatively insensitive to the 
existence of a disk, meaning the solution is very close to that derived for 
single bulges, at least for inclinations less than $1-\cos(i)=0.6$.

The derived bulge-to-disk ratio is very close to the one obtained for single
bulges and disks, for a large range of inclinations and opacities (see
Fig.~\ref{fig:proj_corr_sersic_sersic_bulge_disk_ratios_obs_25}). Only at large
optical depth and does the bulge-to-disk ratio decrease with respect to single
components case. In the K band the corrections are negligible. 

The results presented in
Figs.~\ref{fig:bd_bulge_sersic_disk_sersic_index_difference_obs_25} to
\ref{fig:bd_sersic_sersic_bulge_eff_rad_obs_25} are for $B/D=0.25$. A similar
analysis performed on simulations made with $B/D=0.50$ shows that a more
prominent bulge does not significantly change the results for
$corr^{dust,\,B/D}$ (see examples in Appendix~\ref{sec:app_dust_exp}). As in the case
of fits done with exponential plus S\'{e}rsic
functions, the amplitude of the effects for disks increases with
increasing $B/D$ and decreases for bulges.
Thus,
irrespectively of the fitting functions (exponential plus S\'{e}rsic or 
S\'{e}rsic plus S\'{e}rsic) bulge-disk decompositions of systems containing 
exponential bulges are only a slow varying function of the bulge-to-disk ratio.\\
\begin{figure}[tbh]
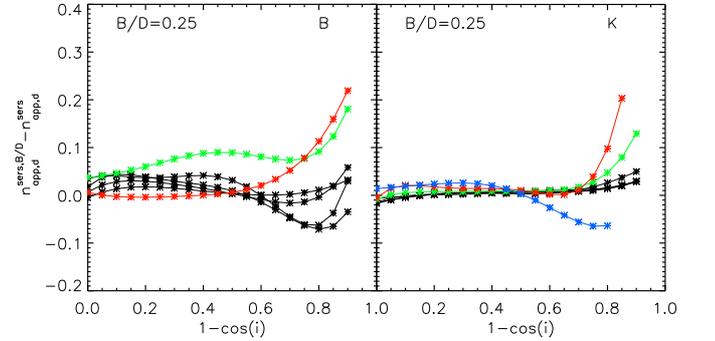
 
 \includegraphics[scale=0.38]{proj_corr_sersic_sersic_bulge_disk_decomp_ndisk_index_variation_obs_25_b.epsi}
 \hspace{-0.33cm}
 \includegraphics[scale=0.38]{proj_corr_sersic_sersic_bulge_disk_decomp_ndisk_index_variation_obs_25_k.epsi}
\caption{\label{fig:bd_bulge_sersic_disk_sersic_index_difference_obs_25} Dust effects
  ($corr^{dust,\,B/D}$) on the derived S\'{e}rsic index of decomposed {\bf disks}, for $B/D=0.25$. 
The solid lines are polynomial fits to the measurements. The plots represent the 
difference between the derived S\'{e}rsic index of decomposed and single 
disks, $n_{app,d}^{sers,B/D}$ and  $n_{app,d}^{sers}$, respectively (corrected
for projection effects), as a
function of inclination ($1-cos(i)$), for the B and K optical bands. 
{\bf Two variable S\'{e}rsic index functions} were used for image decomposition. The black curves are plotted for $\tau_{B}^{f}= 0.1,0.3,0.5,1.0$, 
while the other curves correspond to $\tau_{B}^{f}=2.0$ (green), $4.0$ (red) and $8.0$ (blue).}
\end{figure}
\vspace*{-0.5cm}
\begin{figure}[tbh]
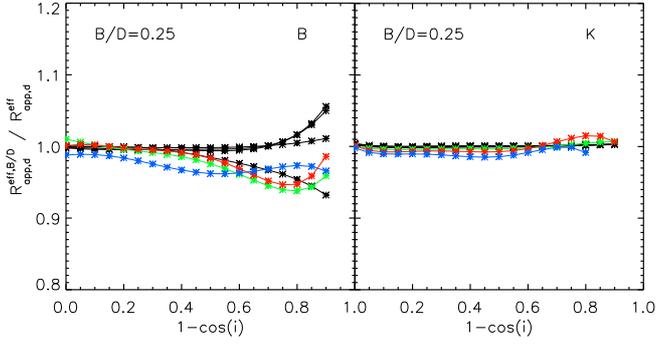

 \includegraphics[scale=0.38]{proj_corr_sersic_sersic_bulge_disk_decomp_scale_length_vs_inclination_obs_25_b.epsi}
 \hspace{-0.33cm}
 \includegraphics[scale=0.38]{proj_corr_sersic_sersic_bulge_disk_decomp_scale_length_vs_inclination_obs_25_k.epsi}
 \caption{\label{fig:bd_sersic_sersic_disk_eff_rad_obs_25} As in
   Fig.~\ref{fig:bd_bulge_sersic_disk_sersic_index_difference_obs_25}, but for 
the derived effective radii of decomposed {\bf disks}.}
\end{figure}
\vspace*{-0.5cm}
\begin{figure}[tbh]
 \includegraphics[scale=0.38]{proj_corr_sersic_sersic_bulge_disk_decomp_nbulge_index_variation_obs_25_b.epsi}
 \hspace{-0.33cm}
 \includegraphics[scale=0.38]{proj_corr_sersic_sersic_bulge_disk_decomp_nbulge_index_variation_obs_25_k.epsi}
\caption{\label{fig:bd_bulge_sersic_sersic_bulge_index_difference_obs_25} 
As in Fig.~\ref{fig:bd_bulge_sersic_disk_sersic_index_difference_obs_25}, but
for the derived S\'{e}rsic index of decomposed {\bf exponential
    bulges}. The effects are
represented as differences between the derived S\'{e}rsic index of decomposed and single 
bulges, $n_{app,b}^{sers,B/D}$ and  $n_{app,b}^{sers}$, respectively.} 
\end{figure}
\begin{figure}[tbh]
 \includegraphics[scale=0.38]{proj_corr_sersic_sersic_bulge_disk_decomp_effective_radius_vs_inclination_obs_25_b.epsi}
 \hspace{-0.33cm}
 \includegraphics[scale=0.38]{proj_corr_sersic_sersic_bulge_disk_decomp_effective_radius_vs_inclination_obs_25_k.epsi}
 \caption{\label{fig:bd_sersic_sersic_bulge_eff_rad_obs_25} 
As in Fig.~\ref{fig:bd_bulge_sersic_disk_sersic_index_difference_obs_25}, but
for the  derived effective radius of decomposed {\bf exponential bulges}.}
\end{figure}
\begin{figure}[tbh]
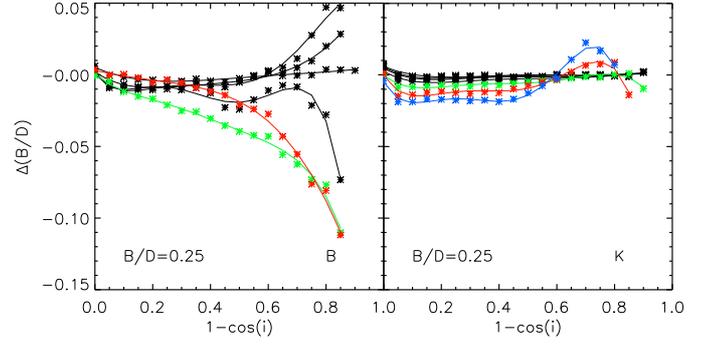

 \includegraphics[scale=0.38]{proj_corr_sersic_sersic_bulge_disk_ratios_obs_25_b.epsi}
\hspace{-0.33cm}
\includegraphics[scale=0.38]{proj_corr_sersic_sersic_bulge_disk_ratios_obs_25_k.epsi}
\caption{\label{fig:proj_corr_sersic_sersic_bulge_disk_ratios_obs_25}
Dust effects ($corr^{dust,\,B/D}$) on the derived bulge-to-disk
ratios, $B/D$. The effects are represented as differences between the apparent
$B/D$ of decomposed disks and bulges as those of single disks and bulges. 
{\bf Two variable S\'{e}rsic index functions} were used for image decomposition. The black curves
are plotted for $\tau_{B}^{f}= 0.1,0.3,0.5,1.0$, while the other curves correspond to $\tau_{B}^{f}=2.0$ (green), $4.0$ (red) and $8.0$ (blue).}
\end{figure}

\subsection{Galaxies with de Vaucouleurs bulges}
\label{subsec:devaucbulges}

In the case of de Vaucouleurs bulges the overall trends are the same as those
for exponential bulges, but with the amplitude of $corr^{dust,\,B/D}$ being
larger for any given inclination and opacity. This means
that for higher S\'{e}rsic indices the decomposition between disk and bulge
starts to be biased. Examples of plots with the corrections are shown in
Appendix~\ref{sec:app_dust_dev}. \\

\section{Single S\'{e}rsic fits}
\label{sec:single}

\begin{figure}[tbh]
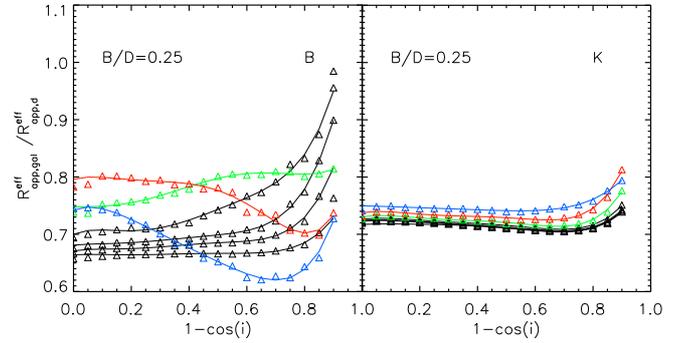

 \includegraphics[scale=0.38]{exp_bulge_disk_single_sersic_eff_radius_vs_disk_eff_radius_obs_25_b.epsi}
 \hspace{-0.33cm}
 \includegraphics[scale=0.38]{exp_bulge_disk_single_sersic_eff_radius_vs_disk_eff_radius_obs_25_k.epsi}
 \caption{\label{fig:bd_exp_bulge_single_sersic_disk_scale_length_ratio_obs_25_b}
Dust effects $corr^{sS}$ on the derived effective radius of galaxies fitted
with {\bf single S\'{e}rsic functions}. The symbols represent the measurements 
while the solid lines are  polynomial fits to the
   measurements. The plots represent the ratio between the effective radius of 
a bulge+disk system and a single disk, $R^{sS}_{app}$ and $R_{app,d}$,
respectively, as a function of inclination ($1-cos(i)$), for the B and K 
optical bands. The black curves are plotted for $\tau_{B}^{f}= 0.1,0.3,0.5,1.0$ (from the bottom towards the top),
while the other curves correspond to $\tau_{B}^{f}=2.0$ (green), $4.0$ (red) and $8.0$ (blue).} 
 \end{figure}

This part of our study is motivated by the fact that single S\'{e}rsic fits 
are commonly used in image analysis (e.g. Hoyos et al. 2011, Simard et
al. 2011, Kelvin et al. 2012, Lackner \& Gunn 2012, Bruce et al. 2012, Bernardi
et al. 2012, H\"au\ss{}ler et al. 2013). This is usually done for large sample of galaxies with marginal
resolution, where morphological components cannot be clearly 
separated/distinguished, or where a two-component fit is not a significant 
improvement over a single S\'{e}rsic fit.

We show here that the derived effective radius of a composite galaxy fitted 
with single S\'{e}rsic functions is strongly underestimated. This can be seen in 
Fig.~\ref{fig:bd_exp_bulge_single_sersic_disk_scale_length_ratio_obs_25_b},
where the effect is visible for both the B and the K band. The strongest
effect appears for the optically thinner cases, where the bulge is biasing
the general solution of the fit. For galaxies with higher optical depth the
attenuation due to dust is flattening the profiles in the centre of the galaxy,
making the effect of bulges less pronounced, and therefore bringing the 
results of single
S\'{e}rsic fits closer to the real size of the disk. The effects strongly
depend on the $B/D$ parameter, with higher values of $B/D$ resulting in a
stronger underestimation of galaxy sizes, for the same inclination and dust
opacity. 

\begin{figure}[htb]
\begin{center}
\includegraphics[scale=0.5]{Simard_disk_size_vs_disk_magn_exp_bulge_red.epsi}
\caption{\label{fig:Simard_disk_size_vs_disk_magn_exp_bulge_red}
Disk size-luminosity relation for a sample of galaxies selected from Simard
et al. (2011). Galaxies with inclinations $1-\cos(i)>0.8$ are
overplotted as red crosses.}
\end{center}
\end{figure}

\section{Application: the inclination dependence of dust effects}
\label{sec:application}

One important application of our modelling is the prediction for the
inclination dependence of the effects of dust on the derived
scale-lengths of disks. To compare our predictions with observations
we used the photometric data derived by Simard et al. (2011) for
galaxies from the Legacy area of the Sloan Digital Sky Survey (SDSS) Data
Release 7. In total Simard et al. performed bulge-disk
decompositions in \textit{g} and \textit{r} bands for 1,123,718 galaxies using three different type of
fits: an exponential disk plus a de Vaucouleurs bulge, an exponential disk plus
a S\'{e}rsic bulge and a single S\'{e}rsic fit. We used the measurements in \textit{r} band for
exponential scale-lengths derived from fits with an exponential
disk plus a S\'{e}rsic bulge. From these we selected only the
measurements for which these fits represent a significant improvement over a single
S\'{e}rsic fit, as listed by Simard et al. We also selected galaxies with redshifts $z \leq
0.08$. This gave us a sample of 117833 galaxies. From this we further
selected galaxies with $B/D<0.35$. This criterion was applied to ensure
a higher probability of selecting a sample of bona fide spiral
disks. This left us with a sample of 38555 galaxies with measured
exponential disk sizes, integrated magnitudes and inclinations. Since
the inclinations listed in Simard et al. (2011) are not corrected for
projection effects (due to the vertical distribution of stars), we
re-calculated these by applying the corrections $corr^{proj}$ from our
model, as listed in Pastrav et al. (2013). In Fig.~\ref{fig:Simard_disk_size_vs_disk_magn_exp_bulge_red} we
show the size-luminosity relation for our sample, as plotted with
black stars. A well defined correlation can be seen, with more
luminous galaxies having larger sizes. The sharp upper bound of the 
distribution is almost certainly due to the surface-brightness limit of the 
photometric imaging
SDSS survey. We also plotted as red
crosses the data corresponding to galaxies with disk inclinations
$1-\cos(i)>0.8$. It is interesting to see that the red points occupy only
the brighter part of the correlation, with most of the points having
disk magnitudes brighter than -17.  No red points exist for the very
faint end of the correlation. This suggests that galaxies with the
smaller axis-ratios are biased towards more luminous galaxies, due to the fact
that low luminosity galaxies with edge-on orientations are missed in the flux
limited survey, which may be
plausibly attributed to the larger attenuation by dust of edge-on
galaxies. We made similar tests for the other bins in inclinations,
where we found no bias. Because of this we excluded the galaxies with
$1-\cos(i)>0.8$ and we only compared the prediction of
our model with data for inclinations in the range $1-cos(i)<0.8$. This
left us with a sample of 33770 galaxies.

To compare our model predictions with the data we derived the average
exponential scale-length for each bin in inclination, where the bins
were taken to be $\Delta\cos(i)=0.05$. For the model predictions we
considered the whole chain of corrections
\begin{figure*}[tbh]
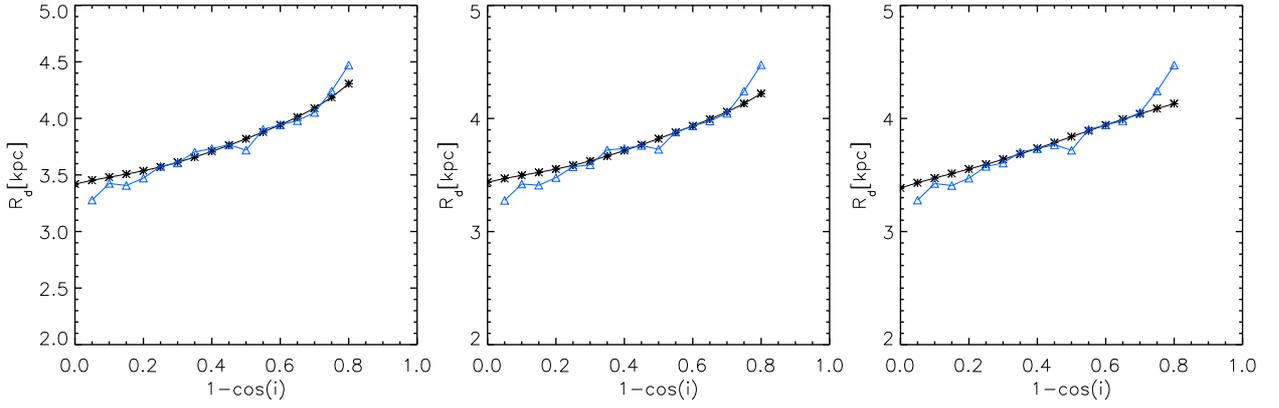

\begin{center}
\includegraphics[scale=0.45]{disk_exp_combined_dust_effects_disk_scalelengths_exp_gal_40_noS_cut_proj_corr_25_b.epsi}
\includegraphics[scale=0.45]{disk_exp_combined_dust_effects_disk_scalelengths_exp_gal_40_noS_cut_25_b.epsi}
\includegraphics[scale=0.45]{disk_exp_combined_dust_effects_disk_scalelengths_exp_gal_40_noS_cut_proj_corr_no_BD_effects_25_b.epsi}
\caption{\label{fig:comparison} {\it Left:} Average inclination dependence of disk sizes for a sample of galaxies
selected from Simard et al. (2011) (blue curve). Overplotted in black
are the predictions of our model for a disk population, scaled to the
averaged disk size derived from the data, at $1-\cos(i)=0.6$,
which corresponds to an intrinsic value of 3.05 kpc. {\it Middle}: The same, but with the projection effects $corr^{proj}$ not included
in the chain of corrections. {\it Right}:  The same, but with decomposition effects $corr^{B/D}$ not included in the chain of corrections.}
\end{center}
\end{figure*}

\begin{eqnarray}
corr(R_d) = corr^{proj}(R_d) * corr^{dust}(R_d) * corr^{B/D}(R_d)
\end{eqnarray}
where $R_d$ is the exponential (radial) scale-length of the stellar
disk, $corr^{proj}(R_d)$ are the projection effects listed in
Pastrav et al. (2013), $corr^{dust}(R_d)$ are the effects
of dust on the scale-length of disks seen in isolation, as listed in
Pastrav et al. (2013),  and $corr^{B/D}(R_d)$ are the effects of dust on
the scale-length of disks seen in combination with a bulge, as derived
in this paper. As in Pastrav et al. (2013), the corrections for an
average population of spiral galaxies were calculated for
$\tau^f_B=4$. The choice for this value of dust opacity was motivated
by the analysis of the attenuation-inclination relation by Driver et
al. (2007), who found an average dust opacity for local universe disk
galaxies of $\tau^f_B=3.8$. A similar average value for comparable
stellar masses was also found by Grootes et al. (2013a). Moreover,
radiative transfer analysis of the UV to FIR SEDs of individual
edge-on galaxies by Misiriotis et al. (2001) and Popescu et al. (2004)
found similar values for $\tau^f_B$.

In the left panel of Fig.~\ref{fig:comparison} we show the comparison of our model
predictions with the data. Overall, the data show the same monotonic
increase in disk sizes with inclination as predicted by our
model. The main contributor to the inclination dependence 
   is due to the effect of dust on the single disk. This can be seen by looking 
  at similar plots, where we omitted the corrections for projection effects
  $corr^{proj}$ (middle panel in Fig.~\ref{fig:comparison}), and the 
  decomposition effects $corr^{B/D}$ (right panel in
  Fig.~\ref{fig:comparison}). In each of the latter cases, the agreement with
  the data is slightly worse. In
  particular the projection and decomposition effects account for the steepening
  of the inclination dependence at higher inclinations. However, the latter
  effects are within the observational errors of the data.

\begin{figure}
\begin{center}
\includegraphics[scale=0.5]{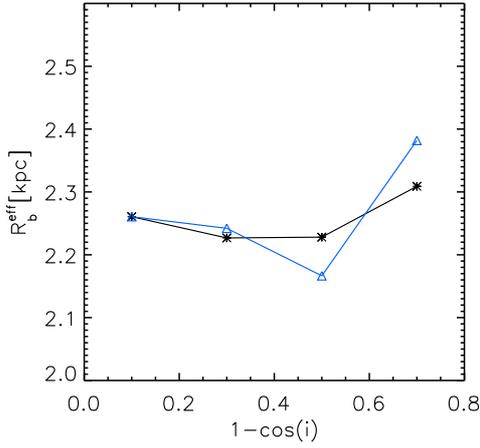}
\caption{\label{fig:dust_BD_bulge_comparison} Average inclination dependence of bulge effective radii for a sample of galaxies
selected from Simard et al. (2011) (blue curve). Overplotted in black
are the predictions of our model for a bulge population, scaled to the
averaged bulge effective radius derived from the data, at 
$1-\cos(i)=[0.0,0.2]$.}
\end{center}
\end{figure}

Using the same sample of $33770$ galaxies considered for the above comparison, 
we used the same approach to study the inclination dependence of bulge effective radii. 
We derived the average bulge effective radius for each bin in inclination (this time considering larger bins with a size of $0.2$ to reduce the noise in the data), while for the model predictions the whole chain of corrections was again considered:
\begin{eqnarray}
corr(R^{eff}_{b}) = corr^{proj}(R^{eff}_{b}) * corr^{dust}(R^{eff}_{b}) * corr^{B/D}(R^{eff}_{b})
\end{eqnarray}
where $R^{ eff}_{b}$ is the effective radius of the stellar bulge.
As before (disks), the model predictions are made for $\tau_B^f=4$.
In Fig.~\ref{fig:dust_BD_bulge_comparison}, we present the result of this 
comparison. Unlike the case of the disks, the data are noisier. 
This may possibly be due to the limited angular resolution of the SDSS data of 
$1.4^{\prime\prime}$ FWHM (Abazajian et al. 2009), which corresponds to 2.3 kpc at the 0.08
redshift limit of the sample considered in this paper.
Nonetheless, both the data and the model show a roughly flat dependence with inclination.
It is interesting to note that  our model predictions for
$\tau^f_B=4$ (considered to be representative for local universe galaxies) can
account for the contrasting inclination dependence of scale-length of disks and
effective radius of bulges.

To conclude, while on average our model can account for
the trends seen in the data, a more detailed analysis of the
inclination dependence of disk and bulge sizes would require both a more
accurate determination of disk scale-lengths and bulge effective radii and an 
analysis done on an object-by-object case. From the point of view of the data, 
a more
accurate determination of sizes would require higher resolution
images, as will soon become available from VISTA/VST. From the point
of view of the analysis, corrections to each data point should be
applied, according to the dust opacity of each galaxy. This, in turn, 
requires determination of $\tau^f_B$. For galaxies with available
panchromatic integrated luminosity densities, determination of $\tau^f_B$ can be
obtained by using the library of radiative transfer model SEDs of
Popescu et al. (2011), the same model that was used to derive the
dust corrections in this paper and in Pastrav et al. (2013). Since the
fits to the SEDs need to be scaled according to the size of the disk,
this becomes an iterative problem to solve. The use of this approach
allows for a self-consistent determination of both intrinsic
parameters of galaxies derived from global measurements and structural
parameters  derived from images. For galaxies without measurements of
integrated dust luminosities, the dust opacity can be derived solely
from optical data, using the method of Grootes et al. (2013a), which
was calibrated by using the same radiative transfer model of Popescu
et al. (2011), again allowing for a self-consistent analysis of both
integrated quantities and structural properties.

\section{Summary}
\label{sec:summary}

In this paper we have presented the results of a study to quantify the effects 
of dust and projection effects
on the derived photometric parameters of disks and bulges obtained from
bulge-disk decomposition. As discussed in Pastrav et al. (2013), these effects
can be separated from the effects of dust and projection effects on single 
disks and bulges. Thus, in  this paper we have only analysed the above
mentioned effects on the decomposition itself. 

We used
simulated images calculated with radiative transfer techniques.
The simulations were produced as part of the large library of dust and PAH
emission SEDs and corresponding dust attenuations presented in Popescu et
al. (2011). All the simulations were calculated using a modified version of the
ray-tracing radiative transfer code of Kylafis \& Bahcall (1987).

We fitted the simulated images with 1D analytic functions available in
GALFIT. The following types of fits were considered: i) fits combining an 
infinitely thin exponential plus a variable-index S\'{e}rsic function for the 
disk and bulge component, respectively, and ii) fits combining 
variable-index S\'{e}rsic functions for both the disk and the bulge. 

The main effects dust has on the bulge-disk decomposition are as
follows:\\\

{\bf Galaxies with exponential bulges}

\begin{itemize}
\item The derived scale-length of a decomposed disk (obtained from fits of type
  i.) is smaller than the derived scale-length of a single disk 
(in the absence of a bulge).
\item The derived axis-ratio of a decomposed disk (obtained from fits of type
  i.) is not changed in the decomposition process.
\item The derived effective radius of a decomposed bulge (obtained from fits of
  type i.) is 
 smaller than the effective radius of a single bulge (in the absence of a 
disk).
\item The derived S\'{e}rsic index of a decomposed bulge (obtained from fits of
  type i.) is slightly smaller than that
  obtained in the absence of a disk.
\item The derived bulge-to-disk ratio (obtained from fits of
  type i.) is smaller that that obtained from single components.
\item The derived effective radius of a decomposed disk (obtained from fits of
  type ii.) is closer to the single disk solution (in
  the absence of a bulge).
\item The derived axis-ratio of a decomposed disk (obtained from fits of type
  ii.) is not changed in the decomposition process.
\item The derived effective radius of a decomposed bulge (obtained from fits
  of type ii.) is
  close to the effective radius  of a single bulge (in the absence
  of a disk).
\item The derived bulge-to-disk ratio (obtained from fits of type
  ii.) is relatively unchanged in the decomposition process.
\item The corrections $corr^{dust,B/D}$ are relatively insensitive to the exact
  value of the $B/D$.
\end{itemize}

{\bf Galaxies with de Vaucouleurs bulges}

\begin{itemize}
\item The overall trends are the same as those
for exponential bulges.
\item The amplitude of the corrections $corr^{dust,B/D}$ is larger than for the 
case  of systems with exponential bulges.
\end{itemize}

The predictions for the inclination dependence of disk scale-lengths and
bulge effective radii were
compared with observational data from a sample selected from Simard et
al. (2011). We show that on average our model can account for the
trends seen in the data. We also show that the main contributor to
the steep increase  of disk scale-length with inclination is the 
effect of dust on single disks, while projection and decomposition effects produce only
secondary effects, strengthening the agreement with the data. We recommend that for more detailed
studies of sizes, an analysis on an object-by-object case should
be involved, in conjunction with determinations of disk
opacities. We show that self-consistent determinations of both intrinsic disk
and bulge sizes
and dust opacities can be obtained using the library of model SEDs of Popescu et
al. (2011) or the method of Grootes et al. (2013a), since these have
been obtained with the same radiative transfer model that was used to
derived the corrections presented in this paper.

\acknowledgements{ We would like to acknowledge the referee for his/her 
very useful comments and suggestions.}

\Online

\begin{appendix}
\section{Examples of projection effects on decomposed disks and bulges for 
galaxies with  exponential bulges and $B/D=0.5$}
\label{sec:app_proj_exp}

\begin{figure}[tbh]
\begin{center}
 \includegraphics[scale=0.7]{proj_eff_scale_length_bd_ratios_intrin_50_b.epsi}
\end{center}
\caption{\textit{Left}: Projection effects on the derived scale-length of decomposed 
   {\bf disks} for $B/D=0.50$. The symbols represent the measurements while 
   the solid lines are polynomial fits to the
   measurements. The plots represent the ratio between the intrinsic 
   scale-lengths of decomposed and single disks, $R^{B/D}_{i,\,d}$ and $R_{i,\,d}$,
   respectively, as a function of inclination ($1-cos(i)$), for the B-band. 
   An {\bf exponential} (disk) {\bf plus a} {\bf variable index S\'{e}rsic} 
   (bulge) distribution were used for image decomposition.
   \textit{Right}: As in the left panel, but for the derived
  bulge-to-disk ratios, $B/D$. The effects are represented as differences 
  between the intrinsic $B/D$ of decomposed disks and bulges and those of
  single disks and bulges.}
\label{fig:proj_disk_scalelength_bd_ratios_50}
\end{figure}
\begin{figure}[tbh]

\begin{center}
 \includegraphics[scale=0.72]{proj_eff_effective_radius_bulge_sersic_index_intrin_50_b.epsi}
\end{center}
\caption{As in Fig.~\ref{fig:proj_disk_scalelength_bd_ratios_50}, but for the
  derived effective radius $R^{eff,\,B/D}_{i,\,b}$  (\textit{Left}) and for the 
  derived S\'{e}rsic indices  (\textit{Right}) of decomposed {\bf exponential
  bulges}. The effects on S\'{e}rsic indices are
   represented as differences between the measured S\'{e}rsic index of 
   decomposed
   and single bulges, $n^{sers,\,B/D}_{i,\,b}$ and $n^{sers}_{i,\,b}$, respectively. }
\label{fig:proj_bulge_effective_radius_sersic_index_50}
\end{figure}

\begin{figure}[tbh]
\begin{center}
 \includegraphics[scale=0.72]{proj_eff_sersic_sersic_eff_radius_disk_sersic_index_intrin_50_b.epsi}
\end{center}
\caption{\textit{Left}: Projection effects on the derived effective radius of 
   decomposed {\bf disks} for $B/D=0.5$. The symbols represent the 
   measurements while the solid lines are polynomial fits to the
   measurements. The plots represent the ratio between the intrinsic 
   effective radius of decomposed and single disks, $R^{eff,\,B/D}_{i,\,d}$ and 
   $R^{eff}_{i,\,d}$, respectively, as a function of inclination ($1-cos(i)$),
   for the B-band. 
   \textbf{Two variable index S\'{e}rsic functions} were used for image decomposition.
   \textit{Right}: As in the left panel, but for the derived
  S\'{e}rsic index, $n^{sers}$. The effects are represented as differences 
  between the measured S\'ersic index of decomposed and single disks, 
  $n^{sers,\,B/D}_{i,\,d}$ and $n^{sers}_{i,\,d}$, respectively.}
\label{fig:proj_disk_effective_radius_sersic_index_50}
\end{figure}

\begin{figure}[tbh]
\begin{center}
 \includegraphics[scale=0.72]{proj_eff_sersic_sersic_eff_radius_sersic_index_intrin_50_b.epsi}
\end{center}
\caption{As in Fig.~\ref{fig:proj_disk_effective_radius_sersic_index_50}, but for
 the decomposed {\bf exponential bulges}.}
\label{fig:proj_bulge_sers_sers_effective_radius_sersic_index_50}
\end{figure}

\begin{figure}[tbh]
\begin{center}
 \includegraphics[scale=0.36]{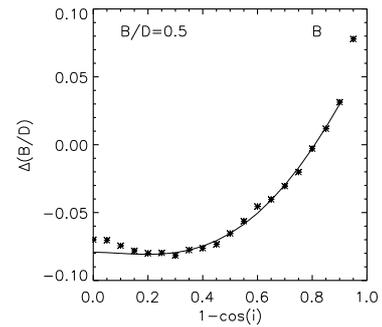}
\caption{As 
in Fig.~\ref{fig:proj_disk_scalelength_bd_ratios_50}, right, but for
  fits done with {\bf two variable S\'{e}rsic index functions}.}
\label{fig:sersic_sersic_bd_ratios_difference_galaxy_intrin_exp_bulge_50_b}
\end{center}
\end{figure}

\clearpage

\section{Examples of projection effects on decomposed disks and bulges for 
galaxies with  de Vaucouleurs bulges and $B/D=0.25$}
\label{sec:app_proj_dev}

\begin{figure}[tbh]
\begin{center}
 \includegraphics[scale=0.7]{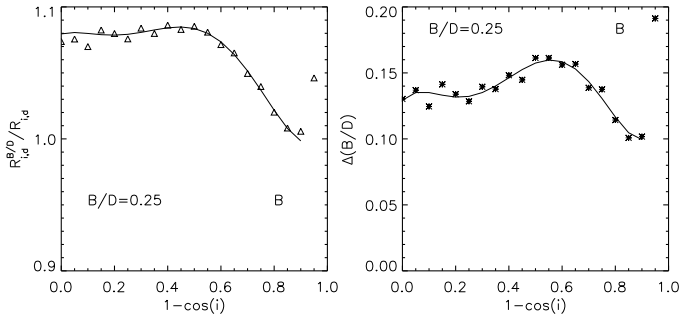}
\end{center}
\caption{\textit{Left}: Projection effects on the derived scale-length of decomposed 
   {\bf disks} for $B/D=0.25$. The symbols represent the measurements while 
   the solid lines are polynomial fits to the
   measurements. The plots represent the ratio between the intrinsic 
   scale-lengths of decomposed and single disks, $R^{B/D}_{i,\,d}$ and $R_{i,\,d}$,
   respectively, as a function of inclination ($1-cos(i)$), for the B-band. 
   An {\bf exponential} (disk) {\bf plus a} {\bf variable index S\'{e}rsic} 
   (bulge) distribution were used for image decomposition.
   \textit{Right}: As in the left panel, but for the derived
  bulge-to-disk ratios, $B/D$. The effects are represented as differences 
  between the intrinsic $B/D$ of decomposed disks and bulges and those of
  single disks and bulges.}
\label{fig:deV_proj_disk_scalelength_bd_ratios_25}
\end{figure}

\begin{figure}[tbh]
\begin{center}
 \includegraphics[scale=0.72]{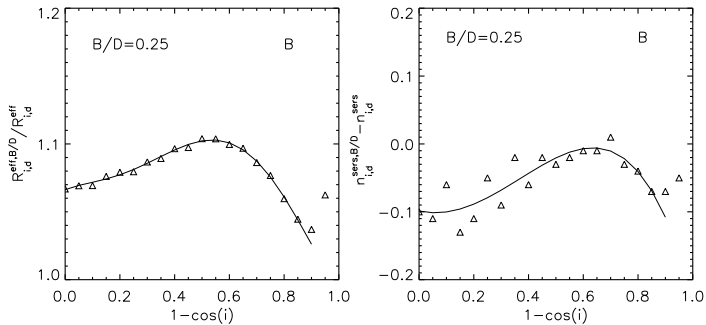}
\end{center}
\caption{\textit{Left}: Projection effects on the derived effective radius of 
   decomposed {\bf disks} for $B/D=0.25$. The symbols represent the 
   measurements while the solid lines are polynomial fits to the
   measurements. The plots represent the ratio between the intrinsic 
   effective radius of decomposed and single disks, $R^{eff,\,B/D}_{i,\,d}$ and 
   $R^{eff}_{i,\,d}$, respectively, as a function of inclination ($1-cos(i)$),
   for the B-band.
   \textbf{Two variable index S\'{e}rsic functions} were used for image decomposition.
   \textit{Right}: As in the left panel, but for the derived
  S\'{e}rsic index, $n^{sers}$. The effects are represented as differences 
  between the measured S\'ersic index of decomposed and single disks, 
  $n^{sers,\,B/D}_{i,\,d}$ and $n^{sers}_{i,\,d}$, respectively.}
\label{fig:deV_proj_disk_effective_radius_sersic_index_25}
\end{figure}
\section{Example of dust effects on decomposed disks and bulges for 
galaxies with exponential bulges and $B/D=0.5$}
\label{sec:app_dust_exp}

\begin{figure}[tbh]
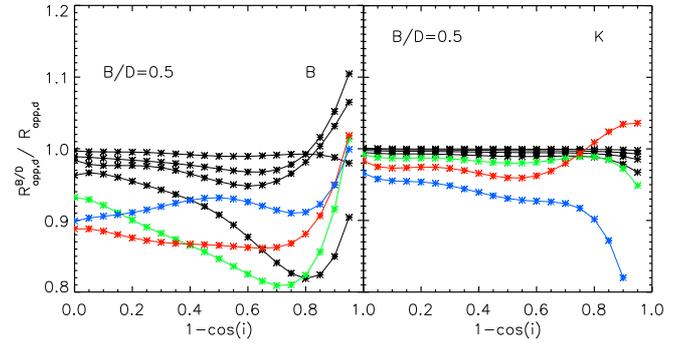

 \includegraphics[scale=0.38]{proj_corr_exp_sersic_bulge_disk_decomp_scale_length_vs_inclination_obs_50_b.epsi}
 \hspace{-0.33cm}
 \includegraphics[scale=0.38]{proj_corr_exp_sersic_bulge_disk_decomp_scale_length_vs_inclination_obs_50_k.epsi}
 \caption{\label{fig:bd_exp_sersic_scale_lengths_obs_50} Dust effects
   ($corr^{B/D}$-$corr^{proj,\,B/D}$) on the derived scale-length of decomposed {\bf disks} for 
   $B/D=0.5$. The solid lines are polynomial fits to the
   measurements. The plots represent the ratio between the apparent 
   scale-lengths of decomposed and single disks, $R^{B/D}_{app,d}$ and $R_{app,d}$,
   respectively, as a function of inclination ($1-cos(i)$), for the B and K 
   optical bands. An {\bf exponential} (disk) {\bf plus a} {\bf variable index 
   S\'{e}rsic} (bulge) distribution were used for image decomposition. The black curves are plotted for $\tau_{B}^{f}= 0.1,0.3,0.5,1.0$,
   while the other curves correspond to $\tau_{B}^{f}=2.0$ (green), $4.0$ (red) and $8.0$ (blue).}
\end{figure}
\begin{figure}[tbh]
 \includegraphics[scale=0.38]{proj_corr_exp_sersic_bulge_disk_decomp_nbulge_index_variation_obs_50_b.epsi}
 \hspace{-0.33cm}
 \includegraphics[scale=0.38]{proj_corr_exp_sersic_bulge_disk_decomp_nbulge_index_variation_obs_50_k.epsi}
\caption{\label{fig:bd_bulge_sersic_index_difference_obs_50} 
As in Fig.~\ref{fig:bd_exp_sersic_scale_lengths_obs_50}, but for the derived 
S\'{e}rsic index of decomposed {\bf exponential bulges}. The effects are
represented as differences between the measured S\'{e}rsic index of decomposed and
single bulges, $n^{sers,\,B/D}_{app,\,b}$ and $n^{sers}_{app,\,b}$, respectively.} 
\end{figure}
\begin{figure}[tbh]
 \includegraphics[scale=0.38]{proj_corr_exp_sersic_bulge_disk_decomp_effective_radius_vs_inclination_obs_50_b.epsi}
 \hspace{-0.33cm}
 \includegraphics[scale=0.38]{proj_corr_exp_sersic_bulge_disk_decomp_effective_radius_vs_inclination_obs_50_k.epsi}
 \caption{\label{fig:bd_exp_sersic_eff_rad_obs_50} 
As in Fig.~\ref{fig:bd_exp_sersic_scale_lengths_obs_50}, but for the derived 
effective radius of decomposed {\bf exponential bulges}.}
\end{figure}
\begin{figure}[htb]
 \includegraphics[scale=0.38]{proj_corr_sersic_sersic_bulge_disk_decomp_scale_length_vs_inclination_obs_50_b.epsi}
 \hspace{-0.33cm}
 \includegraphics[scale=0.38]{proj_corr_sersic_sersic_bulge_disk_decomp_scale_length_vs_inclination_obs_50_k.epsi}
 \caption{\label{fig:bd_sersic_sersic_disk_eff_rad_obs_50} Dust effects
  ($corr^{B/D}$-$corr^{proj,\,B/D}$) on the derived efective radius of
  decomposed {\bf disks}, for $B/D=0.5$. 
   The solid lines are polynomial fits to the measurements. The plots represent the ratio between the apparent 
   effective radius of decomposed and single disks, $R^{eff,\,B/D}_{app,d}$ and $R^{eff}_{app,d}$,
   respectively, as a function of inclination ($1-cos(i)$), for the B and K 
   optical bands.
{\bf Two variable S\'{e}rsic index functions} were used for image
decomposition. The black curves are plotted for $\tau_{B}^{f}=
0.1,0.3,0.5,1.0$, while the other curves correspond to $\tau_{B}^{f}=2.0$ (green), $4.0$ (red) and $8.0$ (blue).}
\end{figure}

\begin{figure}[htb]
 \includegraphics[scale=0.38]{proj_corr_sersic_sersic_bulge_disk_decomp_nbulge_index_variation_obs_50_b.epsi}
 \hspace{-0.33cm}
 \includegraphics[scale=0.38]{proj_corr_sersic_sersic_bulge_disk_decomp_nbulge_index_variation_obs_50_k.epsi}
\caption{\label{fig:bd_bulge_sersic_sersic_bulge_index_difference_obs_50} 
As in Fig.~\ref{fig:bd_sersic_sersic_disk_eff_rad_obs_50}, but
for the derived S\'{e}rsic index of decomposed {\bf exponential bulges}. The effects are
represented as differences between the measured S\'{e}rsic index of decomposed and
single bulges, $n^{sers,\,B/D}_{app,\,b}$ and $n^{sers}_{app,\,b}$, respectively.}
\end{figure}

\begin{figure}[htb]
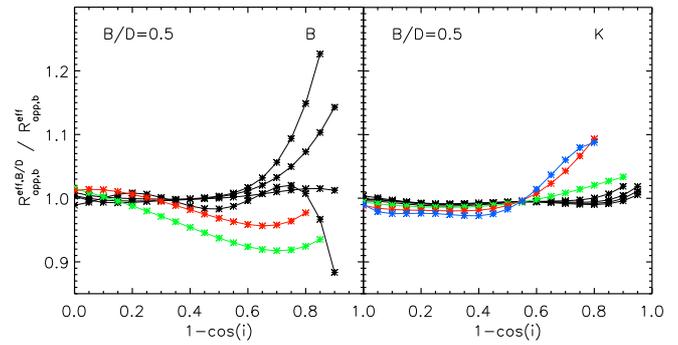

 \includegraphics[scale=0.38]{proj_corr_sersic_sersic_bulge_disk_decomp_effective_radius_vs_inclination_obs_50_b.epsi}
 \hspace{-0.33cm}
 \includegraphics[scale=0.38]{proj_corr_sersic_sersic_bulge_disk_decomp_effective_radius_vs_inclination_obs_50_k.epsi}
 \caption{\label{fig:bd_sersic_sersic_bulge_eff_rad_obs_50} 
As in Fig.~\ref{fig:bd_sersic_sersic_disk_eff_rad_obs_50}, but
for the  derived effective radius of decomposed {\bf exponential bulges}.}
\end{figure}

\clearpage
\section{Examples of dust effects on decomposed disks and bulges for
galaxies with de Vaucouleurs bulges and $B/D=0.25$}
\label{sec:app_dust_dev}

\begin{figure}[htb]
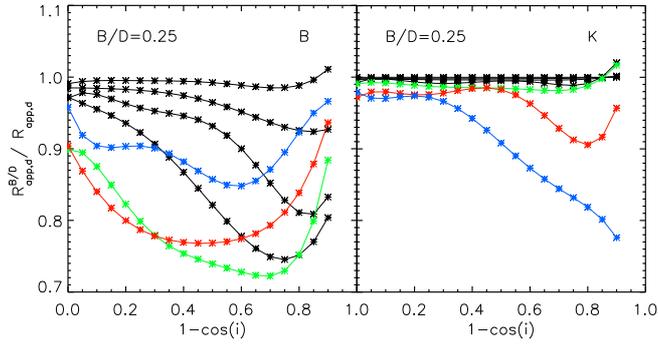

 \includegraphics[scale=0.38]{proj_corr_exp_sersic_bulge4_disk_decomp_scale_length_vs_inclination_obs_25_b.epsi}
 \hspace{-0.33cm}
 \includegraphics[scale=0.38]{proj_corr_exp_sersic_bulge4_disk_decomp_scale_length_vs_inclination_obs_25_k.epsi}
 \caption{\label{fig:bd_exp_sersic4_scale_lengths_obs_25} Dust effects
   ($corr^{B/D}$-$corr^{proj,\,B/D}$) on the derived scale-length of decomposed {\bf disks} for 
   $B/D=0.25$. The solid lines are polynomial fits to the measurements.
   The plots represent the ratio between the apparent 
   scale-lengths of decomposed and single disks, $R^{B/D}_{app,d}$ and $R_{app,d}$,
   respectively, as a function of inclination ($1-cos(i)$), for the B and K 
   optical bands. An {\bf exponential} (disk) {\bf plus a} {\bf variable index 
   S\'{e}rsic} (bulge) distribution were used for image decomposition. 
   The curves are plotted for $\tau_{B}^{f}= 0.1,0.3,0.5,1.0$ 
   (black), $2.0$ (green), $4.0$ (red) and $8.0$ (blue).}
\end{figure}
\begin{figure}[h]
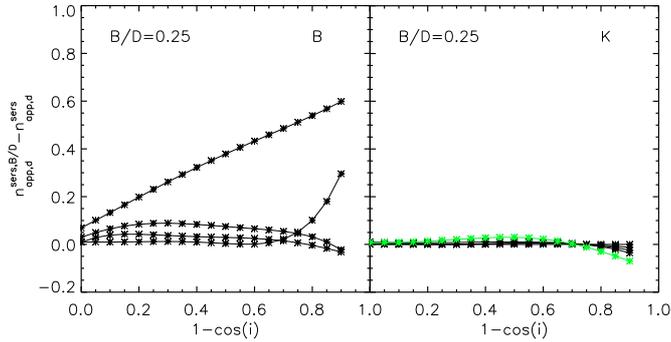
 
 \includegraphics[scale=0.38]{proj_corr_sersic_sersic_bulge4_disk_decomp_ndisk_index_variation_obs_25_b.epsi}
 \hspace{-0.33cm}
 \includegraphics[scale=0.38]{proj_corr_sersic_sersic_bulge4_disk_decomp_ndisk_index_variation_obs_25_k.epsi}
\caption{\label{fig:bd_bulge_sersic_disk_sersic4_index_difference_obs_25} Dust effects
  $corr^{B/D}$ on the derived S\'{e}rsic index of decomposed {\bf disks}, for $B/D=0.25$. 
The solid lines are polynomial fits to the measurements. The plots represent the 
difference between the derived S\'{e}rsic index of decomposed and single 
disks, $n_{app,d}^{sers,B/D}$ and  $n_{app,d}^{sers}$, respectively, as a
function of inclination ($1-cos(i)$), for the B and K optical bands. 
{\bf Two variable S\'{e}rsic index functions} were used for image decomposition. The curves are plotted for $\tau_{B}^{f}= 0.1,0.3,0.5,1.0$ (black), $2.0$ (green) and $4.0$ (red).}
\end{figure}
\begin{figure}[h]
 \includegraphics[scale=0.38]{proj_corr_sersic_sersic_bulge4_disk_decomp_scale_length_vs_inclination_obs_25_b.epsi}
 \hspace{-0.33cm}
 \includegraphics[scale=0.38]{proj_corr_sersic_sersic_bulge4_disk_decomp_scale_length_vs_inclination_obs_25_k.epsi}
 \caption{\label{fig:bd_sersic_sersic4_disk_eff_rad_obs_25} As in Fig.~\ref{fig:bd_bulge_sersic_disk_sersic4_index_difference_obs_25}, but for derived effective radii of decomposed \textbf{disks}.}
\end{figure}
\end{appendix}
\end{document}